\documentclass{article}
\usepackage{epsfig}
\begin{document}
\begin{center}
{\Large\bf STATIC AXIALLY SYMMETRIC SOLUTIONS OF EINSTEIN-YANG-MILLS EQUATIONS
 WITH A NEGATIVE COSMOLOGICAL CONSTANT: THE REGULAR CASE }
\vspace{0.6cm}\\
Eugen Radu
\\
{\small
\it Albert-Ludwigs-Universit\"at, Fakult\"at f\"ur Physik, Hermann-Herder-Stra\ss e 3
\\
 Freiburg D-79104, Germany}
\end{center}

\begin{abstract}
Numerical solutions of the Einstein-Yang-Mills equations 
with a negative cosmological constant are constructed. 
These axially symmetric solutions approach asymptotically the anti-de Sitter spacetime 
and are regular everywhere. 
They are characterized by the winding number $n>1$, the mass and the 
non-Abelian magnetic charge.
The main properties 
of the solutions and the differences with respect to the asymptotically 
flat case are discussed.
The existence of axially symmetric monopole and dyon solutions
in fixed anti-de Sitter spacetime is also discussed.
\end{abstract}
\section{INTRODUCTION}

After the discovery by Bartnik and McKinnon (BK) of a nontrivial particlelike solution of 
the Einstein-Yang-Mills (EYM) equations \cite{bmk}, there has been a great deal of 
numerical and analytical 
work on various aspects of EYM theory.
A large number of self-gravitating 
structures with non-Abelian fields have been found (for a review see \cite{1}).
These include black holes with non-trivial hair, thereby leading to the 
possibility of evading the no-hair conjecture.

Most of these investigations have been carried out on the assumption that spacetime 
is asymptotically flat.
Less is known when the theory is modified to include a cosmological constant 
$\Lambda$ which greatly changes the asymptotic structure 
of spacetime \cite{Hawking}. 

For a positive cosmological constant, the behavior of the solutions
is similar in many 
respects to that of asymptotically flat geometries \cite{cosmBK}.
In particular the
configurations are in both cases unstable \cite{Brodbeck:1996xm}.

If we allow for a negative cosmological constant, 
the solution of the matter free Einstein equations possessing
the maximal number of symmetries is the anti-de Sitter (AdS) spacetime.
Being a maximally symmetric spacetime, it is an excellent 
model to investigate questions of principle related to the quantisation of fields
propagating on a curved background, the interaction with the gravitational field
and issues related to the lack of global hyperbolicity.
Also, lately has been a lot of interest in asymptotically 
AdS spacetimes, connected 
with string theory and related topics.

Recently, some authors have discussed the properties of soliton  and 
black hole solutions of the EYM system for $\Lambda <0$  
($i.e.$ an asymptotically AdS spacetime \cite{4, hoso, 6}). 
They obtained some surprising results, which are strikingly different from the
BK type solutions.
First, there is a continuum of solutions in terms of the adjustable shooting parameter
that specifies the initial conditions at the origin or at the event horizon,  
rather then discrete points.
The spectrum has a finite number of continuous branches.
Secondly there are nontrivial solutions stable against spherically symmetric 
linear perturbations, corresponding 
to stable monopole and dyon configurations.
The solutions are classified 
by non-Abelian electric and magnetic charges 
and the ADM mass.
When the parameter $\Lambda$ approaches zero, 
an already-existing branch of monopoles and dyon solutions 
collapses to a single point in the moduli space.
At the same time new branches of solutions emerge.
A fractal structure in the moduli space has been noticed \cite{hoso, Hosotani:2001iz}.

As observed in \cite{hoso}, these solutions may have profound consequences
in the evolution of the early universe.

In Ref. \cite{8, 9} regular gravitating monopole and dyon solutions in 
Einstein-Yang-Mills-Higgs (EYMH) theory with Higgs field 
in the adjoint representation were shown to exist in asymptotically 
AdS spacetime. 
As happens in asymptotically flat space, a critical value for the Newton 
constant exists above which 
no regular solution can be found.
The presence of a cosmological constant enhances this effect, the critical 
value being smaller than the value found for $\Lambda=0$.

The global existence of a solution of Cauchy problem for the YMH equations in
AdS spacetime is discussed in \cite{Choquet-Bruhat:1989bn}.

However, so far only spherically symmetric solutions have been found. 

For $\Lambda=0$, a SU(2) YM theory coupled 
to a scalar Higgs field \cite{ymh,Hartmann:2000ja},
dilaton \cite{Kleihaus:1997yc} or gravity \cite{kk} is known to possess also
axially symmetric finite energy solutions. 
In the 1990s, the numerical calculation of these configurations 
was one of the most important developments in the domain.  

A  natural question arises: do  non-Abelian axially symmetric solutions
also exist for a nonzero cosmological constant? And if this is the case,
how does the nonasymptotically flat structure 
of spacetime affects these configurations?

The aim of this paper is to address the above questions for the SU(2) gauge group,
under the assumption of axial symmetry, for an asymptotically AdS geometry.

The corresponding problem for a vanishing cosmological constant 
and a purely magnetic gauge field has been 
exhaustively discussed in \cite{kk}.
Representing generalizations of the BK solutions \cite{1},
the solutions obtained by Kleihaus and Kunz in \cite{kk}
have no nonabelian charges  but
are characterized by two integers. 
These are the node number $k$ of the gauge field functions and the winding
number $n$ with respect to the azimuthal angle $\varphi$.
The spherically symmetric BK solutions have winding 
number $n=1$.
A winding number $n>1$ leads to axially symmetric solutions. 
As discussed in \cite{kk}, 
these regular axially symmetric solutions have a torus-like shape.
With the $z-$axis $(\theta=0$) as symmetry axis, the energy density has a strong peak along 
the $\rho$ axis $(\theta=\pi/2$) and decreases monotonically
along the $z-$axis.

The solutions we are looking for are the asymptotically 
AdS analogues of these configurations.

Although  some common features are present,
the results we find are rather different
from those valid in the $\Lambda=0$ case.
A different behavior is noticeable especially for the lower branch of axially 
symmetric solutions.  
These distinctions arise from differences that 
already exist in the spherically symmetric case.

A negative $\Lambda$ implies a continuum of axially symmetric solutions.
For a fixed winding number, they can be classified in a finite 
number of branches and have continuous values of mass and  
non-Abelian magnetic charge.
The radial and angular dependence of the metric and gauge functions can also be different
from the case discussed in \cite{kk}.

The  paper is structured as follows: in the next section we explain the model 
and derive the basic equations, while in 
Section 3 we discuss solutions of YM equations in a 
fixed AdS background.
These solutions were found to be important when discussing
the scaling properties of the mass spectrum \cite{Hosotani:2001iz} 
and have no flat space couterparts.
Some features of the axially symmetric solutions possessing a net YM electric charge
in a fixed AdS background
are also presented in this Section.
The general properties of the axially symmetric gravitating solutions are
 presented in Section 4 where we show results obtained by numerical calculations.
We give our conclusions and remarks in the final section.

\section{GENERAL FRAMEWORK AND EQUATIONS OF MOTION}

\subsection{Einstein-Yang-Mills action}

The basic equations for a static, axially symmetric SU(2) gauge field 
coupled to Einstein gravity (without cosmological term) are well known
(for details see $e. g.$  \cite{kk}).
Here we derive them for a nonzero $\Lambda$, without going into details.
We will follow most of the conventions and notations 
used by Kleihaus and Kunz (KK) in their papers.

The starting point is the Einstein-Yang-Mills action
\begin{equation} \label{lag0}
S=\int d^{4}x\sqrt{-g}[\frac{1}{16\pi G}(\mathcal{R} - 2 \Lambda)
-\frac{1}{2}Tr(F_{\mu \nu }F^{\mu \nu })], 
\end{equation}
where the field strength tensor is
\begin{equation}
F_{\mu \nu} = 
\partial_\mu A_\nu -\partial_\nu A_\mu + i e \left[A_\mu , A_\nu \right] 
\ , \label{fmn} \end{equation}
and the gauge field
\begin{equation}\label{amu}
A_{\mu} = \frac{1}{2} \tau^a A_\mu^a.
\end{equation}
Variation of the action (\ref{lag0})
 with respect to the metric $g^{\mu\nu}$ leads to the Einstein equations
\begin{equation}\label{einstein-eqs}
R_{\mu\nu}-\frac{1}{2}g_{\mu\nu}R +\Lambda g_{\mu\nu}  = 8\pi G T_{\mu\nu},
\end{equation}
where the YM stress-energy tensor is 
\begin{eqnarray}
T_{\mu\nu} = 2{\rm Tr} 
    ( F_{\mu\alpha} F_{\nu\beta} g^{\alpha\beta}
   -\frac{1}{4} g_{\mu\nu} F_{\alpha\beta} F^{\alpha\beta}). 
\end{eqnarray}
Variation with respect to the gauge field $A_\mu$ 
leads to the matter field equations
\begin{equation}\label{YM-eqs}
\nabla_{\mu}F^{\mu\nu}+ie[A_{\mu},F^{\mu\nu}]=0.
\end{equation}
\subsection{Static axially symmetric ansatz and gauge condition}
We generalize 
the isotropic axisymmetric line element considered 
by Kleihaus and Kunz in \cite{kk} for a nonzero $\Lambda$ 
\begin{equation} \label{metric}
ds^2=- f(1-\frac{\Lambda}{3}r^2) dt^2 +  \frac{m}{f} ( \frac{d r^2}
{1-\frac{\Lambda}{3}r^2}+ r^2 d \theta^2 )
           +  \frac{l}{f} r^2 \sin ^2 \theta d\phi^2,
\end{equation}
where the metric functions
$f$, $m$ and $l$ are only functions of 
the coordinates $r$ and $\theta$.

A suitable parametrization of a purely magnetic Yang-Mills connection
 in terms of spherical coordinates is
\begin{eqnarray} \label{ansatz}
A_r=&& \frac{1}{2er} H_{1}(r,\theta) \tau_\phi^n, \nonumber
\\
A_\theta =&& \frac{1}{2e} (1-H_{2}(r,\theta))\tau_\phi^n, \nonumber
\\
A_\phi =&& -n \sin\theta  \left[H_{3}(r,\theta)\frac{\tau_r^n}{2e} +
(1-H_{4}(r,\theta))\frac{\tau_\theta^n}{2e} \right],
\end{eqnarray}
Here the symbols $\tau^n_r$, $\tau^n_\theta$ and $\tau^n_\phi$
denote the dot products of the cartesian vector
of Pauli matrices, $\vec \tau = ( \tau_1, \tau_2, \tau_3) $,
with the spatial unit vectors
\begin{eqnarray}
\vec e_r^{\, n}      &=& 
(\sin \theta \cos n \phi, \sin \theta \sin n \phi, \cos \theta)
\ , \nonumber \\
\vec e_\theta^{\, n} &=& 
(\cos \theta \cos n \phi, \cos \theta \sin n \phi,-\sin \theta)
\ , \nonumber \\
\vec e_\phi^{\, n}   &=& (-\sin n \phi, \cos n \phi,0), 
\end{eqnarray}
respectively.
This ansatz is axially symmetric in the sense that a rotation around the $z-$axis 
can be compensated by a gauge rotation.
It satisfies also some additional discrete symmetries \cite{kk,YMH1}.
To fix the residual abelian gauge invariance we choose the usual
gauge condition \cite{kk}
\begin{equation}\label{gauge}
r \partial_r H_1 - \partial_\theta H_2 = 0. 
\end{equation}
This ansatz contains an integer $n$, representing the winding number with 
respect to the 
azimutal angle $\varphi$.
While $\varphi$ covers the trigonometric circle once, 
the fields wind $n$ times around.
 Note that the spherically symmetric ansatz corresponds to $n=1$, 
$H_1=H_3=0$, $H_2=H_4=w(r)$.
\subsection{Field equations}
From (\ref{einstein-eqs}) and (\ref{YM-eqs}) we obtain a set of seven
nonlinear elliptical partial differential equations which 
can be solved numerically.

Within this specific ansatz and the gauge condition (\ref{gauge}) 
we derive the matter equations
\begin{eqnarray}
0 & = & 
\Bigg( r^2 H_{1,r,r} + H_{1,\theta ,\theta} + H_{2,\theta} 
- n^2 \frac{m}{l} \left( r H_{4,r} H_3 - r H_{3,r} H_4 
+\left( H_3^2 + H_4^2 - 1 \right) H_1 \right)  \Bigg)
 \sin^2 \theta
\nonumber \\  & &
+\left( r H_{2,r} + H_{1, \theta} 
- n^2 \frac{m}{l}\left( 2 H_1 H_3 + r H_{4,r}\right) \right)
\sin \theta \cos \theta- n^2 \frac{m}{l} H_1                    
\nonumber \\  & &
+\sin^2 \theta \left[H_{1, \theta} + r H_{2,r}  \right]  
 \ln \left( \frac{fN\sqrt{l}}{m} \right)_{,\theta} 
+\sin^2 \theta  \left[ r H_{1,r} - H_{2,\theta} \right] r 
\ln \left( \frac{fN\sqrt{l}}{m} \right)_{, r},
\nonumber \\  & &
\label{H1} \\
0 & = & 
\Bigg(r^2 H_{2,r,r} + H_{2,\theta ,\theta} - H_{1,\theta}) 
+n^2 \frac{m}{lN} \left( H_3 H_{4,\theta} -  H_4 H_{3, \theta}
-\left( H_3^2 +H_4^2 - 1 \right) H_2 \right)
\Bigg) \sin^2 \theta               
\nonumber \\  & &
+\left(H_{2,\theta}- r H_{1,r}  
+ n^2 \frac{m}{lN}\left(-2 H_2 H_3 + H_{4,\theta} \right) \right) 
\sin \theta \cos \theta                                                   
+ n^2  \frac{m}{lN}\left( H_4 - H_2 \right) 
\nonumber \\  & &
-\sin^2 \theta \left[ r H_{1,r} - H_{2,\theta}) \right] 
     \ln \left( \frac{fN\sqrt{l}}{m} \right)_{,\theta}
+\sin^2 \theta  \left[ H_{1, \theta} + r H_{2,r}  \right] 
      r \ln \left( \frac{fN\sqrt{l}}{m} \right)_{, r},
\nonumber \\  & &
\label{H2} \\
0 & = & 
\Bigg( r^2 H_{3,r,r} +\frac{1}{N}H_{3,\theta ,\theta} 
- H_3 \left( H_1^2 + \frac{H_2^2}{N} \right) 
+ H_1 \left( H_4 - 2 r H_{4,r} \right) 
- H_4 \left( r H_{1,r} - \frac{H_{2,\theta}}{N} \right)   
\nonumber \\  & &
+\frac{2 H_2 H_{4,\theta} }{N}
\Bigg)  \sin^2 \theta
+\frac{1}{N}\left( H_{3,\theta} + H_2 H_4 - NH_1^2 - H_2^2 
\right) \sin \theta \cos \theta 
-\frac{H_3}{N}
\nonumber \\  & &
+\frac{1}{N}\sin^2 \theta \left[ H_{3,\theta} -1 + H_2 H_4 + \cot \theta H_3  \right] 
\ln \left( \frac{f}{\sqrt{l}}\right)_{,\theta}
 + \sin^2 \theta \left[ r H_{3,r} - H_1 H_4 \right] 
 r \ln \left( \frac{f N}{\sqrt{l}} \right)_{, r},        
\nonumber \\  & &
\label{H3} \\
0 & = & 
\Bigg( r^2 H_{4,r,r} + \frac{1}{N}H_{4,\theta ,\theta} 
-H_4 \left( H_1^2 + \frac{H_2^2}{N} \right) 
- H_1 H_3 
- \frac{H_2}{N} \left( -1 + 2 H_{3,\theta} \right)
+ 2 r H_{3,r} H_1 
\nonumber \\  & &
+H_3 \left( r H_{1,r}-\frac{ H_{2,\theta}}{N} \right) 
 \Bigg) \sin^2 \theta
+\frac{1}{N}\left( -H_2 H_3 - H_1 N + r H_{1,r}N - H_{2,\theta} + H_{4,\theta}  
 \right) \sin \theta \cos \theta
\nonumber \\  & & 
+\frac{H_2 - H_4 }{N}
+\frac{1}{N}\sin^2 \theta \left[ H_{4,\theta} - H_3 H_2 
- \left( H_2 - H_4 \right) \cot \theta \right]
  \ln \left( \frac{f}{\sqrt{l}} \right)_{,\theta}
\nonumber \\  & &
+\sin^2 \theta \left[  r H_{4,r} + H_1 H_3 +  \cot \theta H_1 \right] 
r \ln \left( \frac{fN}{\sqrt{l}} \right)_{, r},            
\label{H4}
\end{eqnarray}
where $N=1-\frac{\Lambda}{3}r^2$.
The resulting equations for the metric functions are
\begin{eqnarray} 
8 \pi G \frac{m}{f} \left( - 2 T_t^t \right)
&=&\frac{1}{r^2} \Bigg( \frac{Nr^2 f_{,r,r}+f_{,\theta ,\theta}+2 Nr f_{,r}}{f} 
-\left( \left( \frac{f_{,\theta}}{f} \right)^2
+ N\left( \frac{r f_{,r}}{f}\right)^2 \right)
+ \cot \theta \frac{f_{,\theta}}{f} 
\nonumber \\  
& &
+\frac{1}{2} \Big( N\frac{r f_{,r}}{f} \frac{r l_{,r}}{l}
  + \frac{f_{,\theta}}{f} \frac{l_{,\theta}}{l} \Big) \Bigg) 
-\frac{\Lambda r}{3}\left ( \frac{l_{,r}}{l}+ 2\frac{f_{,r}}{f}\right) 
+2\Lambda(\frac{m}{f}-1) ,
\label{eqf}
\\
8 \pi G \frac{m}{f} \left( T_r^r +  T_{\phi}^{\phi} \right)
&=&\frac{1}{ 4 r^2} \Bigg(
2 \left( 
\frac{ Nr^2 m_{,r,r} + Nr m_{,r} + m_{,\theta ,\theta}}{m}
-\left( \frac{m_{,\theta }}{m} \right)^2 
-N\left( \frac{ r m_{,r}}{m} \right)^2 
\right) 
+2 \frac{l_{,\theta ,\theta}}{l}  
\nonumber \\
& &  
+ 2 \left( \frac{f_{,\theta}}{f} \right)^2                                                     
- \left( \frac{l_{,\theta}}{l} \right)^2 
+ 2 N\left( \frac{r l_{,r}}{l} + \frac{ r m_{,r}}{m} \right)   
+ N\frac{r m_{,r}}{m} \frac{r l_{,r}}{l}
- \frac{ m_{,\theta}}{m} \frac{ l_{,\theta}}{l}
\nonumber \\ 
&&
-2 \left( \frac{m_{,\theta}}{m} -2 \frac{l_{,\theta}}{l} 
\right) \cot \theta 
\Bigg) 
-\frac{\Lambda r}{6}\left ( \frac{l_{,r}}{l}+ 2\frac{m_{,r}}{m}\right) 
+2\Lambda(\frac{m}{f}-1), 
\\
8 \pi G \frac{m}{f} \left( T_r^r 
+  T_{\theta}^{\theta} \right)
&=&\frac{1}{4 r^2} \Bigg(
 2  \frac{Nr^2 l_{,r,r}+l_{,\theta ,\theta}}{l} 
 -\left( \frac{l_{,\theta}}{l} \right)^2
 -N\left( \frac{r l_{,r}}{l} \right)^2 
 + 6 N\frac{r l_{,r}}{l} + 4 \frac{l_{,\theta}}{l} \cot \theta \Bigg)
\nonumber \\ 
&&
-\frac{\Lambda rl_{,r}}{2l}+2\Lambda(\frac{m}{f}-1).
\label{eqm3} 
\end{eqnarray}
In the above relations, the components of the energy-momentum tensor are
\begin{eqnarray}
T_r^r
&=&
\frac{1}{2}\left(\frac{f}{m}\right)^2\frac{N}{r^2}F_{r \theta}^2
+\frac{1}{2}\frac{f^2N}{mlr^2 \sin^2 \theta }F_{r \phi}^2
-\frac{1}{2}\frac{f^2}{mlr^4 \sin^2 \theta }F_{\theta \phi}^2,
\nonumber\\
T_{\theta}^{\theta}
&=&
\frac{1}{2}\left(\frac{f}{m}\right)^2\frac{N}{r^2}F_{r \theta}^2
-\frac{1}{2}\frac{f^2N}{mlr^2 \sin^2 \theta }F_{r \phi}^2
+\frac{1}{2}\frac{f^2}{mlr^4 \sin^2 \theta }F_{\theta \phi}^2,
\nonumber\\
T_{\phi}^{\phi}
&=&
-\frac{1}{2}\left(\frac{f}{m}\right)^2\frac{N}{r^2}F_{r \theta}^2
+\frac{1}{2}\frac{f^2N}{mlr^2 \sin^2 \theta }F_{r \phi}^2
+\frac{1}{2}\frac{f^2}{mlr^4 \sin^2 \theta }F_{\theta \phi}^2,
\nonumber\\
-T_{t}^{t}
&=&
\frac{1}{2}\left(\frac{f}{m}\right)^2\frac{N}{r^2}F_{r \theta}^2
+\frac{1}{2}\frac{f^2N}{mlr^2 \sin^2 \theta }F_{r \phi}^2
+\frac{1}{2}\frac{f^2}{mlr^4 \sin^2 \theta }F_{\theta \phi}^2,
\nonumber
\end{eqnarray}
where, similar to \cite{kk} we define
\begin{eqnarray}
F_{r \theta}^2
& = & \frac{1}{r^2}
\left[ (H_{1,\theta}+rH_{2,r})^2+(rH_{1,r}-H_{2,\theta})^2\right],
\nonumber\\
F_{r \phi}^2
& = & \frac{n^2 \sin^2 \theta}{r^2}
\left[ (rH_{3,r}-H_1H_4)^2+(rH_{4,r}+H_1H_3+\cot\theta H_1)^2\right],
\nonumber\\
F_{\theta \phi}^2
& = & n^2 \sin^2 \theta
\left[ \left( H_{4,\theta}-H_2H_3-\cot \theta (H_2-H_4)\right )^2
   +(H_{3,\theta}-1+H_2H_4+\cot\theta H_3)^2\right].
\nonumber
\end{eqnarray}
\subsection{Boundary conditions}
To obtain  asymptotically AdS regular solutions
with finite energy density
the metric functions have to satisfy the boundary conditions 
\begin{equation} \label{b1} 
\partial_r f|_{r=0}= \partial_r m|_{r=0}= \partial_r l|_{r=0}= 0,
\end{equation}
\begin{equation}\label{b2}
f|_{r=\infty}= m|_{r=\infty}= l|_{r=\infty}=1.
\end{equation}
We assume also the flatness condition \cite{kramer}
\begin{equation}
m|_{\theta=0}=l|_{\theta=0}.
\end{equation}
The boundary conditions satisfied by the matter functions are
\begin{equation}\label{b3}
H_{2}|_{r=0}=H_{4}|_{r=0}= 1, \ \ \ H_{1}|_{r=0}=H_{3}|_{r=0}=0,
\end{equation}
 at the origin and 
\begin{equation}\label{b4}
H_2|_{r=\infty}=H_4|_{r=\infty}= \omega_0, \ \ \ 
H_1|_{r=\infty}=H_3|_{r=\infty}=0,
\end{equation}
at infinity, where $w_0$ is an undetermined constant.
For a solution with parity reflection symmetry, 
the boundary conditions along the axes are
\begin{equation}\label{b5}
\begin{array}{lllllllllll}
H_1|_{\theta=0,\pi/2}&=&H_3|_{\theta=0,\pi/2}&=&0,
\\
\partial_\theta H_2|_{\theta=0,\pi/2} 
&=& \partial_\theta H_4|_{\theta=0,\pi/2}
&=& 0,
\end{array}
\end{equation}
\begin{equation}\label{b6}
\begin{array}{lllllll}
\partial_\theta f|_{\theta=0,\pi/2} 
&=& \partial_\theta m|_{\theta=0,\pi/2} &=&
\partial_\theta l|_{\theta=0,\pi/2} &=&0. 
\end{array}
\end{equation}
Therefore we need to consider the solutions only in the region $0\leq \theta \leq \frac{\pi}{2}$. 
Regularity on the $z-$axis requires 
\begin{equation}
H_2|_{\theta=0}=H_4|_{\theta=0}.
\end{equation}
Dimensionless partial differential equations are obtained by the following rescaling
$r \to (\sqrt{4\pi G}/e) r,~\Lambda \to (e^2/4 \pi G) \Lambda$.
In  Bjoraker-Hosotani conventions, this corresponds to taking 
a unit value for the parameter $v= 4\pi G/e^2$ \cite{hoso}.
\subsection{The mass and the Yang-Mills charges of the solution}

At spatial infinity, the line element (\ref{metric}) can be written as
\begin{equation}\label{mass1}
ds^2=ds_0^2+h_{\mu \nu}dx^{\mu \nu},
\end{equation}
where $h_{\mu \nu}$ are deviations from the background AdS metric $ds_0^2$.
Similar to the asymptotically flat case, one expects 
the values of conserved quantities to be encoded in the functions $h_{\mu \nu}$.
The construction of these quantities for an asymptotically AdS spacetime 
was addressed for the first time in the eighties (see for instance 
Ref. \cite{Abbott:1982ff, Henneaux:1985tv}).
However, the generalization of Komar's formula in this case is not 
straightforward and 
requires the further subtraction of a background configuration in order 
to render a finite result. 

Using the Hamiltonian formalism, Henneaux and Teitelboim \cite{Henneaux:1985tv}
have computed the mass of an asymptotically AdS spacetime in the following way.
They showed that the Hamiltonian must be supplemented by surface terms
in order to be consistent with the equations of motion.
These surface terms yield conserved charges associated with the Killing vectors
of an asymptotic AdS geometry.
The general expression of a conserved quantity is
\begin{equation}\label{mass2}
J_A=\frac{1}{16 \pi}
\oint d^2 S_i [G^{ijkl}(\xi_{A}^{\bot}
{\stackrel{\circ}{\nabla}}_{j}g_{kl}
-h_{kl}{\stackrel{\circ}{\nabla}}_{j}\xi_{A}^{\bot})+
2\xi_A^k \pi_k^i],
\end{equation}
where $G^{ijkl}=\frac{1}{2}(-{\stackrel{\circ}{g}})^{1/2}
({\stackrel{\circ}{g}}^{ik}{\stackrel{\circ}{g}}^{jl}
+{\stackrel{\circ}{g}}^{il}{\stackrel{\circ}{g}}^{jk}
-2{\stackrel{\circ}{g}}^{ij}{\stackrel{\circ}{g}}^{kl})$,
$\xi_{A}^{\bot}$ is the component of the Killing vector
$\xi_{A}$ in the direction of the unit normal to the hypersurface $t=const.$,
$h_{ik}$ is the deviation from the AdS metric,
$\pi_k^i$ is the canonical momentum and ${\stackrel{\circ}{\nabla}}_j$
is the covariant derivative with respect tot the background three-metric
${\stackrel{\circ}{g}}_{ik}$.
The total energy is the charge associated with the Killing vector
 $\partial/\partial t$. 

It has been also checked that the Henneaux-Teitelboim  energy,
the Brown-York energy \cite{Brown:1993br} and the 
Einstein and Landau-Lifshitz energy \cite{Abbott:1982ff}
all agree for the Kerr-AdS spacetime.
However, for the same spacetime, 
the generalized Komar mass has a different value \cite{Pinto-Neto:1995ci}.

In this work we use the Henneaux-Teitelboim formalism to compute numerically the 
mass-energy of a gravitating EYM configuration.
Substituting the ansatz (\ref{metric}) in (\ref{mass2}) 
yields the following expression for the total mass
\begin{equation}\label{mass3}
M=\lim_{r \rightarrow \infty} \frac{\Lambda}{12}
\int_{0}^{\pi/2} d \theta \sin \theta 
\left(-\frac{m-l}{r f}+\frac{\partial}{\partial r}(\frac{m+l}{f})\right)r^4.
\end{equation}
The dimensionless mass is obtained by using the rescaling 
$M \to (eG/\sqrt{4\pi G}) M$.

Solutions of the field equations are also classified by the
nonabelian electric and magnetic charges $Q_E$ and $Q_M$.  
The nonabelian charges defined by
\begin{equation}\label{charge}
\pmatrix{Q_E\cr Q_M\cr}
= {e\over 4\pi} \int dS_k \, \sqrt{-g} \, 
Tr \pmatrix{ F^{k0}\cr \tilde F^{k0} \cr} \tau_r
\end{equation}
are conserved because the Gauss flux theorem \cite{hoso}. 

Within our ansatz $Q_M=n(1-\omega_0^2)$, $Q_E=0$.

\section{YANG-MILLS FIELDS IN FIXED ADS BACKGROUND}
Because the asymptotic structure of geometry is different, 
in an AdS spacetime one does not have to couple the YM system to scalar 
fields or gravity in order to obtain finite energy solutions.
Here the cosmological constant breaks the scale 
invariance of pure YM theory to give finite energy solutions.

It is the purpose of this section to present both analytical and numerical
arguments for the existence
of nontrivial monopoles and dyons solutions of pure YM equations in a 
four-dimensional AdS spacetime, gravity being regarded as a fixed
field ($i.e.$ $f=l=m=1$; also, in this section we don't use the
 above discussed rescaling).

Although being extremely simple, nevertheless this model appears to contain all 
the essential features of the Bjoraker-Hosotani solutions.
In this way, what might seem surprising ($e.g.$ the existence of stable 
solutions) finds a natural explanation.

The existence of these nongravitating 
solutions has  recently been noticed in \cite{Hosotani:2001iz}
when discussing the scaling behavior of the EYM monopoles and dyons.

\subsection{Spherically symmetric solutions}
We start by briefly discussing the solutions obtained for $n=1$ 
within the ansatz (\ref{ansatz}). 
In this case, the equation of motion has the simple form
\begin{eqnarray} \label{weq}
(\omega'(1-\frac{\Lambda}{3}r^2))'&=&\frac{\omega(\omega^2-1)}{r^2},
\end{eqnarray}
where the prime denotes derivative with respect to $r$.
The numerical results show the existence of a one-parameter
family of solutions regular at $r=0$ with the behavior
familiar from the gravitating case
\begin{eqnarray}\label{origin}
w(r) &=& 1-br^2 +O(r^4) ,
\end{eqnarray}
where $b$ is an arbitrary constant.
The asymptotic expansion at large $r$ is
\begin{eqnarray}\label{infinity}
w = w_0+w_1\frac{1}{r} + \cdots,
\end{eqnarray}
where $w_0$, $w_1$ are constants to
be determined by numerical calculations.
These boundary conditions permit a non-vanishing magnetic charge $Q_M$.

The overall picture we find 
 is rather similar to 
the one described in \cite{hoso} where gravity is taken into account.
By varying the parameter $b$, a continuum of monopole solutions is obtained.
As a new feature, we notice the existence of zero- and one-node monopole 
solutions ($k=0$, $1$) only.
Also, for a fixed value of $\Lambda$, we obtain finite energy solutions 
for only
one interval in parameter space, $b_{min}<b<b_{max}$ (with $b_{min}<0$). 
The energy of the solutions is an increasing function of the absolut value of $b$ and diverges at 
the extremities of the interval. 
The allowed values of $b$ corresponds approximately 
to the lower branch found in \cite{hoso} when coupling to gravity.
This is not an unexpected feature. 
We recall that, given a flat spacetime soliton solution,
one can expect that it will have (asymptotically flat) gravitating generalizations.
Apart from the fundamental gravitating solutions
a sequence of radial excitations is likely to exist \cite{1, maison}.
For example, in flat spacetime
the YMH system (with a doublet scalar field) has only one-node solutions; 
when gravity is included, the solutions exist for all $k$ \cite{gmo}.

For large enough $r$, it is possible to write
an approximate expression of $w(r)$ in terms of elliptic functions.
A nontrivial exact solution of the YM equations is
\begin{eqnarray} \label{exact}
\omega=1/(1-\Lambda/3 r^2)^{1/2},
\end{eqnarray} 
describing a monopole in AdS spacetime with unit magnetic charge 
and mass $\sqrt{(-3\Lambda)} \pi/8e^2$.
It should be mentioned that this is not a new result.
The existence of this exact solution has been noticed 
for the first time in \cite{Boutaleb-Joutei:1979va}
for a positive cosmological constant and a different coordinate system. 

The arguments presented in \cite{hoso} for the (linear) stability of the 
nodeless $n=1$ monopole 
solutions apply directly to the nongravitating case.
In Bjoraker-Hosotani analysis (Ref. \cite{hoso}, Section VII), this corresponds to taking 
 $v= 0,~H=1-\Lambda r^2/3,~p=1$, while $\delta H= \delta p=\delta m=0$.
We starts with the most general expression of a spherically symmetric YM connection 
\begin{eqnarray}\label{YM-a-new}
A=\frac{1}{2e} \, \Big\{ u(r,~t)\tau_3dt + \nu(r,~t) \tau_3 dr
+ (w(r,~t)\tau_1+\tilde{w}(r,~t)\tau_2)d\theta
\nonumber
\\
+(\cot\theta\tau_3+w(r,~t)\tau_2-\tilde{w}(r,~t)\tau_1)\sin\theta d\phi \Big\}.
\end{eqnarray}
All field variables are written as the sum of the static 
equilibrium solution whose stability we are investigating
and a time dependent perturbation.
In examining time-dependent fluctuations around finite energy solutions it is
convenient to work in the $\delta u(r,t)=0$ gauge. 
By following the standard methods
we derive linearized equations for $\delta w(r,t)$, $\delta \tilde
w(r,t)$ and $\delta \nu(r,t)$. 
The linearized equations for $\delta w(r,t)$ and $\delta \nu(r,t)$ 
are obtained in \cite{hoso} for the more general gravitating case
(as usual $\delta \tilde w(r,t)$ is determined by $\delta \nu(r,t)$).
For a harmonic time dependence $e^{i \Omega t}$, the linearized system
implies two standard Schr\"odinger equations.
The analysis of the potential's properties
in these Schr\"odinger equations can be done following Ref. \cite{hoso}.
The standard arguments presented by Bjoraker and Hosotani  
are still valid and imply that for nodeless solutions
there are no negative eigenvalues for $\Omega^2$ and thus no unstable modes.

\subsection{Axially symmetric solutions}
In addition to these spherically symmetric solutions we study
their axially symmetric generalizations.
Subject to the  boundary conditions (\ref{b1})-(\ref{b6}),
we solve the YM equations numerically.

Our methods are similar
to those used by the authors of \cite{kk} in their works. 
The KK scheme solves the  field equations
following an iteration procedure.
One starts with a known spherically 
symmetric configuration and increases
the winding number in small steps.
The field equations are first discretized on a 
nonequidistant grid and the resulting system
is solved iteratively until convergence is achieved.
In this scheme, a new radial variable is introduced
which maps the semi-infinite region $[0,\infty)$ to the closed region $[0,1]$.
Thus the region of integration is not truncated 
and the model converges to a higher accuracy.
There are various possibilities for this transformation,
but a choice which is flexible enough was $x = \frac{r}{c+r}$,
 where $c$ is a properly chosen constant.
Typical grids have sizes $170 \times 30$, 
covering the integration region 
$0\leq{x}\leq 1$ and $0\leq\theta\leq\pi/2$.

The numerical calculations are performed by using the program
FIDISOL, based on the iterative Newton-Raphson
method. Detailes on the FIDISOL code are presented in \cite{FIDISOL}.
To obtain  axially symmetric solutions, 
we start with the $n=1$ solution discussed above
 as initial guess and increase the value of $n$ slowly.
The iterations converge, and repeating the procedure one obtains
in this way solutions for arbitrary $n$.
The physical values of $n$ are integers.
The numerical error for the functions is estimated to be 
lower than $10^{-3}$. 

The energy density of these nongravitating
solutions is given by the
$tt$-component of the energy momentum tensor $T_{\mu}^{\nu}$; integration
over all space yields their mass-energy
\begin{equation}\label{mass}
E = \int{\left\{\frac{1}{4} F_{ij}^a F_{ij}^a+\frac{1}{2} F_{it}^a F_{it}^a 
\right\} \sqrt{-g}d^3x} \ 
=2\pi \int_0^{\infty}r^2dr\int_0^{\pi}\sin\theta d\theta 
 {\left\{\frac{1}{4} F_{ij}^a F_{ij}^a+\frac{1}{2} F_{it}^a F_{it}^a 
\right\} } .
\end{equation}  
The energy and the magnetic charge of solutions are proportional
to their winding number $n$.

For every spherically symmetric configuration ($i.e.$ a given value of parameter $b$) 
we have obtained higher winding number generalizations. 
Moreover, the branch structure noticed for $n=1$ 
is retained for higher winding number solutions.
These solutions have very similar properties with 
the corresponding EYM counterparts.
Therefore, the general picture we present here applies also in the next section.


In Fig. 1 we present the gauge functions $H_i$ 
and the energy density $\epsilon$ as function of the radial coordinate
$r$ for the angles $\theta=0$, $\pi/4$ and $\pi/2$ for three different solutions.
Here the winding number is $n=3$ and $\Lambda=-0.01$; 
also the mass is given in units $4 \pi /e^2$.
The configuration with ($Q_M=3$, $M=0.332$) represents a 
higher winding generalization of the 
exact solution (\ref{exact})($i.e.~ b=0.00166$ and $w_0=0$);
we suspect the existence of a general  analytic form of 
this solution (valid for $n\geq 1$).

The configurations with $\omega_0 \neq 0$ have been arbitrarily selected;
the solution with 
($Q_M=2.391$, $M=0.576$) is obtained starting with a spherically
symmetric solution with $b=0.003$, while for 
($Q_M=-13.284$, $M=1.118$) we have $b=-0.001$.

The functions $H_2$ and $H_4$ have a small $\theta$ dependence,  
although the angular dependence of 
matter functions generally increases with $Q_M$.
We notice also that the gauge field function $H_1$ 
remains nodeless and for every solution with $w_0>1$ it takes only negative values
($H_1$ and $H_3$ are zero on the axes in Fig.~1a and 1c).

In Fig.~2 we show the energy density $\epsilon$
and the gauge functions $H_i$
for a nodeless solution with $n=3$, $k=0$, total mass $M=5.481$ (in units $4 \pi /e^2$), 
magnetic charge 
$Q_{M}=-37.185$ 
as a function of the
compactified coordinates $\rho = x \sin \theta$ and
$z = x \cos \theta$ (for $\Lambda=-0.01$) (here $x=r/(50+r)$).
The parameter $b$ for the corresponding spherically symmetric solution is $b=-0.0015$.
In this case, the functions 
$H_2$ and $H_4$ are almost spherical.
The function $H_1$ does not possess a non-trivial node 
and takes only negative  values.

\subsection{Nongravitating dyon solutions}
The existence of dyon solutions without a Higgs field
is a new feature for AdS spacetime \cite{hoso}.
If $\Lambda \ge 0$ the electric part of the gauge fields
is forbidden \cite{hoso,bizon}. In order for the
boundary conditions at infinity to permit the electric 
fields and maintain a finite ADM mass 
we have to add scalar fields to the theory. 
\newline
The YM ansatz (\ref{ansatz}) can be generalized to include an electric part 
(see $e. g.$ \cite{Hartmann:2000ja})
\begin{eqnarray} \label{electric}
A_t = H_{5}(r,\theta) \frac{\tau_r^n}{2e} +
H_{6}(r,\theta) \frac{\tau_\theta^n}{2e} .
\end{eqnarray}
A possible set of boundary conditions for the electric potentials $H_5, H_6$ is
\begin{eqnarray} 
H_{5}|_{r=0}=H_{6}|_{r=0}=0,\ \ H_5|_{r=\infty}=u_0,\ \ H_{6}|_{r=\infty}=0,
\end{eqnarray}
and 
\begin{equation}
\partial_\theta H_5|_{\theta=0,\pi/2}=H_6|_{\theta=0,\pi/2}=0
\end{equation}
for a solution with parity reflection symmetry.
The equations of motion in this case are
\begin{eqnarray}
0 & = & 
\Bigg( r^2 H_{1,r,r} + H_{1,\theta ,\theta} + H_{2,\theta} 
- n^2 \left( r H_{4,r} H_3 - r H_{3,r} H_4 
+\left( H_3^2 + H_4^2 - 1 \right) H_1 \right)  \Bigg)
 \sin^2 \theta
\nonumber \\  & &
+\left( r H_{2,r} + H_{1, \theta} 
- n^2 \left( 2 H_1 H_3 + r H_{4,r}\right) \right)
\sin \theta \cos \theta- n^2 H_1                    
\nonumber \\  & &
+\sin^2 \theta \left[ r H_{1,r} - H_{2,\theta} \right] r 
\ln \left( N \right)_{, r}
+\frac{\sin^2 \theta r^2}{N}\left[rH_{5,r} H_6-rH_{6,r} H_5+H_1(H_5^2+H_6^2)\right],
\nonumber \\  & &
\label{dH1} \\
0 & = & 
\Bigg(r^2 H_{2,r,r} + H_{2,\theta ,\theta} - H_{1,\theta}) 
+ \frac{n^2}{N} \left(H_3 H_{4, \theta} - H_4 H_{3,\theta}   
-\left( H_3^2 +H_4^2 - 1 \right) H_2 \right)
\Bigg) \sin^2 \theta               
\nonumber \\  & &
+\left(H_{2,\theta} - r H_{1,r}  
+ \frac{n^2}{N}\left( -2 H_2 H_3 + H_{4,\theta} \right) \right) 
\sin \theta \cos \theta                                                   
+ \frac{n^2}{N}\left( H_4 - H_2 \right) 
\nonumber \\  & &
+\sin^2 \theta  \left[ H_{1, \theta} + r H_{2,r}  \right] r \ln \left( N \right)_{, r}
+\frac{\sin^2 \theta r^2}{N^2}\left[H_5 H_{6,\theta}-H_{6} H_{5,\theta}+H_2(H_5^2+H_6^2)\right],
\nonumber \\  & &
\label{dH2} \\
0 & = & 
\Bigg( r^2 H_{3,r,r} +\frac{1}{N}H_{3,\theta ,\theta} 
- H_3 \left( H_1^2 + \frac{H_2^2}{N} \right) 
+ H_1 \left( H_4 - 2 r H_{4,r} \right) 
- H_4 \left( r H_{1,r} - \frac{H_{2,\theta}}{N} \right)   
\nonumber \\  & &
+\frac{2 H_2 H_{4,\theta} }{N}
\Bigg)  \sin^2 \theta
+\frac{1}{N}\left( H_{3,\theta} + H_2 H_4 - NH_1^2 - H_2^2 
\right) \sin \theta \cos \theta 
-\frac{H_3}{N}
\nonumber \\  & &
 + \sin^2 \theta \left[ r H_{3,r} - H_1 H_4 \right] r \ln \left( N \right)_{, r} 
+ \frac{\sin^2 \theta r^2 H_6}{N^2} \left[ H_4H_5 +H_6(H_3+\cot \theta) \right],         
\nonumber \\  & &
\label{dH3} \\
0 & = & 
\Bigg( r^2 H_{4,r,r} + \frac{1}{N}H_{4,\theta ,\theta} 
-H_4 \left( H_1^2 + \frac{H_2^2}{N} \right) 
- H_1 H_3 
+ \frac{H_2}{N} \left(1 - 2 H_{3,\theta} \right)
+ 2 r H_{3,r} H_1 
\nonumber \\  & &
+H_3 \left( r H_{1,r}-\frac{ H_{2,\theta}}{N} \right) 
 \Bigg) \sin^2 \theta
+\frac{1}{N}\left( -H_2 H_3 -N H_1  + r H_{1,r}N - H_{2,\theta} + H_{4,\theta}  
 \right) \sin \theta \cos \theta+ \frac{H_2 - H_4}{N}
\nonumber \\  & & 
+\sin^2 \theta \left[  r H_{4,r} + H_1 H_3 +  \cot \theta H_1 \right] 
r \ln \left( N \right)_{, r} 
+ \frac{\sin^2 \theta r^2 H_5}{N^2} \left[ H_4 H_5 +H_6(H_3+\cot \theta) \right] ,           
\nonumber \\  & &
\label{dH4}\\
0 & = & 
\Bigg( r^2 H_{5,r,r} + \frac{1}{N}H_{5,\theta ,\theta} 
+2rH_{6,r}H_1+ rH_{1,r}H_6-\frac{1}{N}\Big(
2H_{6,\theta}H_2+H_6(H_{2,\theta}+n^2 H_3H_4)
\nonumber \\  & & 
+H_5(NH_1^2+H_2^2+n^2H_4^2)\Big)\Bigg)\sin^2 \theta
+\frac{1}{N}\left(H_{5,\theta}-H_6(H_2+n^2H_4)\right)
\sin \theta \cos \theta,
\nonumber \\  & &
\label{dH5}\\
0 & = & 
\Bigg( r^2 H_{6,r,r} + \frac{1}{N}H_{6,\theta ,\theta} 
-rH_{1,r}H_5-2rH_{5,r}H_1+\frac{1}{N}
\Big(2H_{5,\theta}H_2+H_5(H_{2,\theta}-n^2H_3H_4)-H_6(NH_1^2
\nonumber \\  & & 
+H_2^2
+n^2H_3^2-1)\Big)\sin^2 \theta
-\frac{n^2 H_6}{N}+\frac{1}{N}
\Big(H_{6,\theta}+H_2H_5+n^2(2H_3H_6+H_4H_5)\Big)
\sin \theta \cos \theta.
\end{eqnarray}
Similar to the asymptotically flat case \cite{wald},
a vanishing $u_0$ implies a purely magnetic solution.
To prove this, we express the electric part in (\ref{mass})
\begin{eqnarray} \label{electric-mass1}
E_e =\frac{1}{2} \int{ F_{it}^a F_{it}^a \sqrt{-g}d^3x}
\end{eqnarray}
as a surface integral at infinity.
We use also the existence of the Killing vector $\partial/ \partial t$ which implies
\begin{eqnarray} 
F_{i t} = D_{i}A_t 
\end{eqnarray}
and the YM equations (\ref{YM-eqs}). 
Thus we obtain the general result
\begin{eqnarray} \label{electric-mass2}
-E_e &=&Tr( \int \{ D_{i}(A_t F^{i t} \sqrt{-g}) -
A_t D_{i}(F^{it} \sqrt{-g}) d^3x \})
\end{eqnarray}
and, for a regular configuration
\begin{eqnarray}
E_e &=& Tr( \oint_{\infty}A_t F^{r t} dS_{r} ).
\end{eqnarray}
Therefore, for our ansatz
\begin{eqnarray}
E_e &=& \frac{\pi u_0}{e^2} \lim_{r \rightarrow \infty} 
\int_{0}^{\pi/2} r^2 \partial_{r} H_5 \sin \theta d \theta .
\end{eqnarray}
This result provides also an useful test to verify the accuracy 
of the numerical calculations.

When taking $n=1$, $H_5=u(r)$ and $H_6=0$ we find spherically 
symmetric nongravitating dyon solutions.
In this case we have
\begin{eqnarray}\label{origin-e}
u(r) &=& ar+\frac{a}{5}( -2b +\frac{1}{3}\Lambda ) r^3+O(r^5),
\end{eqnarray}
at the origin (where $a$ and $b$ are arbitrary constants) and 
\begin{eqnarray}\label{infinity-e}
u &=& u_0+u_1\frac{1}{r} + \cdots
\end{eqnarray}
at large $r$, where $u_0$, $u_1$, $w_0$, $w_1$ are constants to
be determined by numerical calculations.
The expansion for $w(r)$ is still valid.
These boundary conditions permit non-vanishing charges $Q_M$ and $Q_E$.
In order to obtain the value of these charges
at some distance, we have calculated 
the integrand in (\ref{charge}) in the numerical code.

If the shooting parameter $a$ is nonzero, we find dyon solutions. 
Solutions are found for a continuous set of parameters $a$ and $b$; 
for some limiting values of these parameters solutions blow up.
Given $(a, ~b$) the general behavior of the gauge functions $w, u$ 
is similar to 
the gravitating case; there are also solutions with $Q_M=0$ but $Q_E \ne 0$.
The surprising numerically 
properties noticed for the gravitating case \cite{hoso} are found also for our solutions 
(for example, 
$Q_M \simeq -1/\sqrt{4 \pi}$ at $b=0.0061$ independent of the value of $a$).
In fact we suspect that these properties reside in the nongravitating sector of the theory.

When studying dyon solutions we notice the existence 
of higher node ($k>1$) configurations. Also, 
there are solutions where $w$ does not cross the $r$ axis. 
For a fixed value of $b$, the number of nodes is determined by the value of the parameter $a$.
Typical spherically symmetric 
solutions are displayed in Fig. 3.

The same numerical method described above is used to obtain
higher winding number dyon solutions.

Axially symmetric dyon solutions are also 
known to exist in a SU(2) Yang-Mills-Higgs 
(when working in flat spacetime background).
In the Prasad-Sommerfeld limit, they are known analytically, 
while for finite Higgs self-coupling
they have been recently constructed numerically \cite{Hartmann:2000ja}.
To the author's knowledge there are no known regular axially symmetric
 dyon solutions (analytical or numerical) in a nonflat geometry. 
The toroidal shape of the energy density of the monopole solutions is retained 
for the $n>1$ dyon solutions, as is illustrated in Fig. 4g.
As a typical axially symmetrical configuration, we show 
in this three-dimensional plot 
the gauge functions $H_i$
and the energy density $\epsilon$
for the a solution with $n=2$, $k=1$, total energy $E=1.046$ (in units $4 \pi /e^2$), 
nonabelian charges 
$Q_{M}=-4.124$ and $Q_{E}=-4.362$ 
as a function of the compactified coordinates $\rho = x \sin \theta$ and
$z = x \cos \theta$ (for a better visualization we define here $x=r/(100+r)$).
The value of the cosmological constant is $\Lambda=-0.01$.
This solution has been obtained starting from a spherically symmetric 
configurations with the shooting parameters $b=0.01$ and $a=0.005$.
As seen in Fig. 4a-f, 
 the gauge functions $H_2, ~H_4, ~H_5$
do not exhibit a strong angular dependence. 
The $H_1$, $H_3$ functions remain nodeless
while the second electric potential $H_6$ always
presents  a complicated nodal structure and angular dependence.

The effect of the presence of the YM electric charge is seen in Fig.~5,
where the energy density is shown as a function of
the radial coordinate for several values of
the angle $\theta$, both for an axially symmetric dyon solution and a
monopole solution with the same values of 
magnetic charge and winding number $(\Lambda=-0.01)$.

\section{AXIALLY SYMMETRIC SOLUTIONS IN THE PRESENCE OF GRAVITY}
In their paper \cite{hoso} Bjoraker and Hosotani have used 
Schwarzschild-like coordinates with a line element
\begin{equation}\label{metric3}
ds^2=- \frac{H(\tilde r)}{p(\tilde r)^2}dt^2 +  \frac{1}{H(\tilde r)} d \tilde r^2 
  + \tilde r^2 \left( d \theta^2 + \sin^2 \theta d\phi^2 \right),
\end{equation}
where 
\begin{equation}\label{H}
H(\tilde r) = 1 - \frac{2 \tilde m(\tilde r)}{\tilde r}-
\frac{\Lambda}{3} \tilde r^2 .
\end{equation}
Since we want to use the spherically symmetric solution
as the starting point for the calculation of axially symmetric
configurations, it is appropriate to
transform the known Bjoraker-Hosotani solution to the coordinates which 
appear in our general metric (\ref{metric}).

Given the presence of a cosmological constant,
this coordinate transformation is more complicated 
then the transformation in \cite{kk}.

\subsection{Coordinate transformation}

By requiring $l=m$ and the metric functions  $f$ and $m$
to be only functions of the coordinate $r$,
the axially symmetric isotropic metric (\ref{metric})
reduces to the form
\begin{equation}\label{metric4}
ds^2= -f(1-\frac{\Lambda}{3}r^2) dt^2 
+ \frac{m}{f}  (\frac{dr^2}{1-\frac{\Lambda}{3}r^2} + r^2 \left( d \theta^2 
+\sin^2 \theta d\phi^2 \right) ).  
\end{equation}
The relations (\ref{metric3})-(\ref{metric4}) yields
\begin{equation}\label{co1}
\frac{f(r)}{m(r)} = \frac{r^2}{\tilde r^2}=\beta^2
\end{equation}
and
\begin{equation}\label{r-tilda}
\frac{dr}{r\sqrt{1-\frac{\Lambda }{3}r^2}} =
\frac{1}{\sqrt{H(\tilde r)}} \frac{d \tilde r}{\tilde r}.
\end{equation}

Since the mass function $\tilde m(\tilde r)$ is only known numerically,
we have to numerically integrate (\ref{r-tilda}) 
to obtain $r(\tilde r)$.
Therefore we find
\begin{equation}\label{co5}
\beta(\tilde r)=\frac{2\Psi}{1+\frac{\Lambda}{3}\tilde r^2 \Psi^2},
\end{equation}
with
\begin{equation}\label{co6}
\Psi(\tilde r) =  \exp \left[ { -\int_{\tilde r}^\infty
\frac{1}{\tilde r'} \left( \frac{1}{\sqrt{H}}-1 \right) d \tilde r' }
\right]
\end{equation}
The integration constant is adjusted such that at infinity
$\beta =1$, i.e.~$r= \tilde r$.
The integrand in (\ref{co6}) is well behaved at the origin, since
$m(\tilde r) \sim \tilde r^3$ \cite{hoso}.

For a first branch solution, 
the values of the metric functions are close to one.
Fig.~6a demonstrates the coordinate transformation
for higher branch spherically symmetric
solutions with $k=1-3$ and different magnetic charges. 
The metric functions $f,~m$
and the gauge field function $w$ are shown in figure Fig.~6b.
These solutions resemble those obtained for $\Lambda=0$,
 with quantitative differences only.
For example a smaller value of metric function $m$ at the origin has to be noticed.

\subsection{Numerical method}
We employ the same numerical algorithm as for the YM solutions in fixed AdS background 
presented above.
To obtain  axially symmetric solutions, 
we start with a $n=1$ EYM solution as initial 
guess and increase the value of $n$ slowly 
(for a fixed $\omega_0$).
A second procedure, also employed for the first branch solutions, 
is to start with a known axially symmetric YM solution (with $n=2,3,...$)
as an initial guess for the full system.

The numerical error for the functions is estimated to be 
on the order of $10^{-3}$ or lower for first branch solutions and $10^{-2}$ 
in rest.
This error depends also on the magnetic charge and mass of the solutions. 
Axially symmetric generalizations of the $n=1$ solutions with a large mass 
are difficult to obtain. 

A set of $\Lambda=0$ test runs was carried out, 
primarily designed to evaluate the code's ability to reproduce the KK results.
In this case, we have obtained an excellent agreement with the results of
\cite{kk}.

\subsection{Properties of the solutions}
For all the solutions we present 
we take a cosmological constant $\Lambda=-0.01$,
also the main value considered in Ref.~\cite{hoso}.
However, a similar general behavior has been found 
for other negative values of $\Lambda$.

Starting from a spherically symmetric configuration 
we obtain higher winding number generalizations 
with many similar properties.
For a fixed winding number, the solutions can also be indexed 
in a finite number of branches classified
by the mass and the non-Abelian magnetic charge.
This is in sharp contrast to the $\Lambda=0$ case,
where only a discrete set of solutions is found \cite{kk}.

The metric functions $f,~m,~l$ are completely regular 
and show no sign of an apparent horizon.

We begin with a description of the lowest branch
axially symmetric regular solutions.
In this case the winding number is $n>1$ and  
nodeless or one-node solutions are allowed. 
These solutions are of particular interest because they are likely 
to be stable against linear perturbations (for $k=0$). 
As expected, the gauge functions $H_{i}$ 
looks very similar to those of the 
corresponding (pure-) YM solutions.
The general picture presented in Fig.~2
is valid in this case too.
The typical values of the metric functions  $m, ~f, ~l$ 
are closed to one. These functions
do not exhibit a strong angular dependence,
while $m$ and $l$ have a rather similar shape.

To see the change of the functions 
for an increasing $n$,
we exhibit in Fig.~7 first branch solutions with $k=1$,~$\omega_0=-3$, 
and $n=1$, $2$ and $3$. 
In Figs.~7a-d the gauge field functions are shown,
in Figs.~7e-g the metric functions,
and in Fig.~7h the energy density of the matter fields.
These two-dimensional plots exhibit the $r$ dependence
for three fixed angles $\theta=0$, $\pi/4$ and $\pi/2$.
Note that the $H_1$, $H_3$ functions remain nodeless
($H_1$ and $H_3$ are zero on the axes in Figs.~7a,~c as required by the boundary conditions (\ref{b5})).
As expected, the angular dependence of the metric and matter functions
increases with $n$.
However, the location of the nodes of the gauge field functions
$H_2$, $H_4$ does not move further outward with increasing $n$.
At the same time the peak
of the energy density along the $r$-axis slightly  
shifts outward with increasing $n$ and increases in height.
At the origin the values of the metric functions decreases with $n$.

This behavior contrasts with the picture obtained in an asymptotically
flat spacetime.

In Fig. 8 the mass of the first branch solutions $M$
is plotted as a function of the nonabelian magnetic charge $Q_M$ 
for various winding numbers.
For the studied configurations we find that the 
total mass of a gravitating solution has a smaller value than the mass of the
corresponding solution in a fixed AdS background (for fixed $Q_M,~\Lambda$).
This inequality is of course in accord with our intuition
that gravity tends to reduce the mass. A similar property has been noticed 
for monopole solutions 
in a spontaneously broken gauge theory (without cosmological constant) \cite{Lee}.

Beside these fundamental gravitating solutions, EYM theory possesses
also excited solutions not presented in a fixed AdS background.

In this case, the metric functions of this EYM solution are
considerably smaller at the origin, and the gauge field functions
have their peaks and nodes shifted inwards,
as compared to the first branch solutions.
The energy density of the matter fields has higher peaks, 
which are shifted inwards, compared to the fundamental
gravitating configurations.
Otherwise, many properties of the branch axially symmetric solutions
are similar to those of their asymptotically flat counterparts.

To see the change of the functions for a second branch solution,
we exhibit in Figs.~9 a three-dimensional plot for a configuration with 
winding number $n=2$, node number $k=1$, magnetic charge 
$Q_M=1.18$
and total mass $M=1.498$.
Fig.~9h presents the energy density of the matter fields $\epsilon$,
showing a pronounced peak along the $\rho$-axis
and decreasing monotonically along the $z$-axis (here $x=r/(1+r)$.
Equal density contours presented here reveal a torus-like shape 
of the solutions.

The same general behavior is obtained for two-node solutions.

We don't address in this paper the problem of limiting solutions,
which is still unclear even in the spherically symmetric case.
Using the metric form (\ref{metric3}) Bjoraker and Hosotani
have observed that, as the parameter $b$ (in (\ref{origin}))
is increased, the function $H(\tilde{r})$ hits
zero from above for some values of $\tilde r$.
Also, when $b=b_c$, $k$ has a finite value,  
$\omega(\tilde{r}_h)=p(\tilde{r}_h)=H'(\tilde{r}_h)=0$
and the space ends at $r=\tilde{r}_h$.
There is a universality in the behavior of the critical
 solutions \cite{hoso}.
The meaning of the critical spacetime is yet to be clarified.

This behavior strongly contrast with the asymptotically flat case.
There are no restrictions on the node number $k$ and, 
as $k \to \infty$, the BK solutions tend to a configuration that is 
the union of two parts.
A non-trivial part for $\tilde r<1$
represents an oscillating solution, and a simple part
for $\tilde r>1$, represents the exterior of an extremal 
Reissner-Nordstr\o m (RN) solution with mass $M=1$ and charge 
$Q_M=1$ \cite{limit}.
The limiting axially symmetric configuration represents 
the exterior of an extremal RN solution with mass $n$ and 
charge $Q_M=n$ \cite{kk}.

We have found difficult it to obtain axially symmetric 
generalizations of the spherically
symmetric solutions near the critical spacetime,
with large errors for the functions.
A different metric parametrization appears to be necessary.

\section{CONCLUDING REMARKS}
In this paper we have presented numerical arguments that
EYM theory with a negative cosmological constant possesses
regular static axially symmetric solutions.
They generalize to higher winding number 
the known spherically symmetric solutions.

We started by presenting arguments that SU(2)-YM theory possess
solutions with nonvanishing magnetic and 
electric charges and arbitrary winding number
when AdS spacetime replaces the Minkowski 
space as the ground state of the theory.
The spherically symmetric solutions 
we found have  properties similar to the lower branch of
their known gravitating counterparts.

When including gravity,
we have presented results suggesting  the existence of axially symmetric solutions.
These configurations have continuous values of mass and non-Abelian 
magnetic charges and present a branch structure.
As seen from the figures,
the distributions of the mass-energy density 
$-T_{t}^{t}$ can be different from those of spherical configurations
($i.e.$ almost toroidal distributions).
As a result of these distributions of mass-energy density,
 the spacetime structure
of our solutions can be considerably nonspherical, strongly axisymmetric.
We have noticed a somewhat different behavior 
of the $k=1$ fundamental gravitating solutions as compared to higher nodes excitations. 

We have not  considered  the question of stability for higher winding number solutions.
However, we expect that the nodeless, lower branch of axially symmetric
solutions is stable.
A rigorous proof is however desirable, 
analogous to the proof given for the spherically symmetric case.

In Ref. \cite{Hosotani:2001iz}  a scaling law is derived  for the mass spectrum of the 
spherically symmetric solutions 
with respect to their nonabelian charges $Q_E$ and $Q_M$, $\Lambda$ and the parameter
$v= 4\pi G/e^2$.
The mass of monopoles and dyons is expressed in terms of a universal
function $f(Q_M, Q_E)$ independent on $v,~\Lambda$ and also on $k$. 
The monopole and dyon solutions  in the lowest branch ($k=0$) are
essentially the  solutions in the fixed AdS background metric and will be stable.  
The solutions in the higher branches ($k>0$) are obtained by dressing monopole 
and dyon solutions in the fixed AdS background metric around the
BK solutions in the asymptotically  flat space.
As all BK solutions are unstable, the monopole and dyon solutions in
the higher branches are also unstable.

We suspect the existence of a similar behavior 
for the axially symmetric monopole solutions discussed in this paper.
We have found already that for $n>1$ the lowest branch monopole solutions are 
essentially solutions in a pure YM theory.
Here the BK solutions are replaced with
the KK solutions.
The axial symmetry will introduce a new parameter to 
the universal scaling function, the winding number $n$. 

As discussed in \cite{hoso, Hosotani:2001iz}, the soliton solutions depend nontrivially
on the value of $\Lambda$.
A fractal structure in the moduli space of the solutions 
has been observed \cite{hoso, Matinian:2000zq}. 
New branches emerge as $|\Lambda|$ becomes smaller
and the shape of branches has approximate self-similarity.
As $\Lambda \to 0$ solutions on a branch collapse
to a point - the BK solution, and the nodeless solutions disappear as their ADM mass vanishes.
We didn't investigate this point in the present work, restricting ourselves
to $\Lambda=-0.01$.
However, it is very probable that this general picture remains 
valid for the axially
symmetric configurations and the continuum of solutions 
becomes a discrete set as  $\Lambda \to 0$.
Here also, in this limit, the BK solutions 
are replaced with the KK generalizations.

Actually, due to conformal
invariance of the YM equations and the fact that the AdS metric is reduced to
 the flat metric
by a conformal transformation, any solution on AdS background corresponds to
some solution in the flat space.
In Minkowski coordinates these are nonstationary solutions.
For example, the monopole solution (\ref{exact}) corresponds to 
the  well-known flat-space meron solution \cite{meron}.

Analogous solutions with higher winding number should also exist when 
including 
a Higgs or a dilaton  field in a theory with $\Lambda<0$.
We conjecture the existence of axially symmetric  
gravitating YM black hole solutions
with a nonvanishing cosmological constant.
These would be the AdS spacetime generalizations of the asymptotically flat 
solutions discussed in \cite{Kleihaus:1998ws}.
The axially symmetric nodeless solutions are of particular interest,
because they are likely to be stable 
against linear perturbations.
For $\Lambda>0$, axially symmetric 
EYM solutions generalising 
the spherically symmetric configurations 
found in \cite{cosmBK} should exist also.

Finally, this is not the complete story: 
the investigations can be extended
to the gravitating axially symmetric dyon solutions.
This can be an important issue, since in the asymptotically flat case
this problem has not been solved yet.
For $\Lambda \geq 0$, there are no-go theorems forbidding the
spherically symmetric dyon  regular solutions \cite{bizon, hoso}.
Also, the authors of \cite{wald} 
conjectured the absence of charged regular EYM solutions.
However, as shown in \cite{Brodbeck:1997ek} the BK solutions admit slowly 
rotating charged generalizations. The total angular momentum of these solutions
is proportional to the non-Abelian electric charge.
Therefore it is natural to look for (gravitating-) axially symmetric dyon
solutions, which however have not been found 
so far within a nonperturbative approach.

In a theory with a negative cosmological constant,
the boundary conditions at infinity allow
for spherically symmetric dyon solutions.
Starting from the spherically symmetric $n=1$ solutions, 
we have obtained 
nongravitating axially symmetric dyon YM configurations.
We suppose that these configurations will survive 
when coupling to gravity.
\\
\\
{\bf Acknowledgments}

The author is grateful to Professor J.J. van der Bij 
for helpful advice and encouragement.
A useful discussion with Professor Jutta Kunz is also acknowledged.

 This work was performed in the context of the
Graduiertenkolleg of the Deutsche Forschungsgemeinschaft (DFG):
Nichtlineare Differentialgleichungen: Modellierung,Theorie, Numerik, Visualisierung.



%
%
%
%
%
%
%
%
%
%
%
%
\newpage
{\bf Figure Captions}
\newline
Figure 1:
The gauge functions $H_i$
and the energy density $\epsilon$
(in units $4 \pi /e^2$)  are shown as a function of the radial coordinate
$r$ for the angles $\theta=0$, $\pi/4$ and $\pi/2$ for three nongravitating solutions
with ($Q_M=2.391$, $M=0.576$), ($Q_M=3$, $M=0.332$) and ($Q_M=-13.284$, $M=1.118$).
Here the winding number is $n=3$ and $\Lambda=-0.01$.
\newline

Figure 2:
The gauge functions $H_i$ and the energy density $\epsilon$
(in units $4 \pi /e^2$) 
for a monopole solution with $n=3,~M=5.481,~Q_M=-37.185$
are  shown as a function of the compactified coordinates $z, \rho$.
\newline

Figure 3:
Typical nongravitating spherically symmetric solutions for $\Lambda=-0.01$,  
a fixed value of the parameter $b=0.001$ 
and $a=0.,~0.01,~0.02$. 
The figure for $a=0$ corresponds to a magnetic monopole.
The energy density $\epsilon(r)$ is given in units  $4 \pi /e^2$.
\newline

Figure 4:
The gauge functions $H_i$ 
and the energy density $\epsilon$ (in units $4 \pi /e^2$) for a nongravitating 
dyon solution with 
$n=2,~k=1,~E=1.046$,~$Q_M=-4.124,~Q_E=-4.36$,
are  shown as a function of the compactified coordinates $z, \rho$.
\newline

Figure 5:
The energy density $\epsilon$ (in units $4 \pi /e^2$) 
for a dyon solution with $E=1.116$,
$~Q_M=-2.272,~Q_E=6.43$ and for a
monopole solution with $E=0.607$ and the same value of $Q_M$  
is shown for several values of the angle $\theta$ ($n=2$).
\newline

Figure 6(a):
The coordinate transformation between the isotropic coordinate $r$
and the Schwarzschild-like coordinate $\tilde r$
is shown for spherically symmetric solutions with $k=1-3$.
\newline

Figure 6(b):
The metric functions $f,~l$ and the gauge field function $w$
are shown as a function of the the isotropic coordinate $r$
for the solutions presented in Fig. 6a.
\newline

Figure 7:
The metric functions $f$, $l$, $m$, the gauge functions $H_i$
and the mass density are shown as a function of the radial coordinate
$r$ for the  angles $\theta=0$, $\pi/4$ and $\pi/2$.
Here $n=1$,~2, and 3, $k=1$, $Q_M/n=-8$, $M(n=1)=0.955$, $M(n=2)=2.351$ and  $M(n=3)=4.179$.
\newline

Figure 8:
Mass $M$ is plotted as a function of magnetic charge $Q_M$
for first branch gravitating monopole solutions at $\Lambda=-0.01$.
The winding number $n$ is also  marked.
\newline

Figure 9:
The metric functions $f$, $l$, $m$, the gauge functions $H_i$
and the mass density $\epsilon$ for a monopole solution with $n=2,~M=1.498, 
~Q_M=1.18$,
are  shown as a function of the compactified coordinates $z, \rho$.

\newpage

\setlength{\unitlength}{1cm}

\begin{picture}(16,16)
\centering
\put(-2,0){\epsfig{file=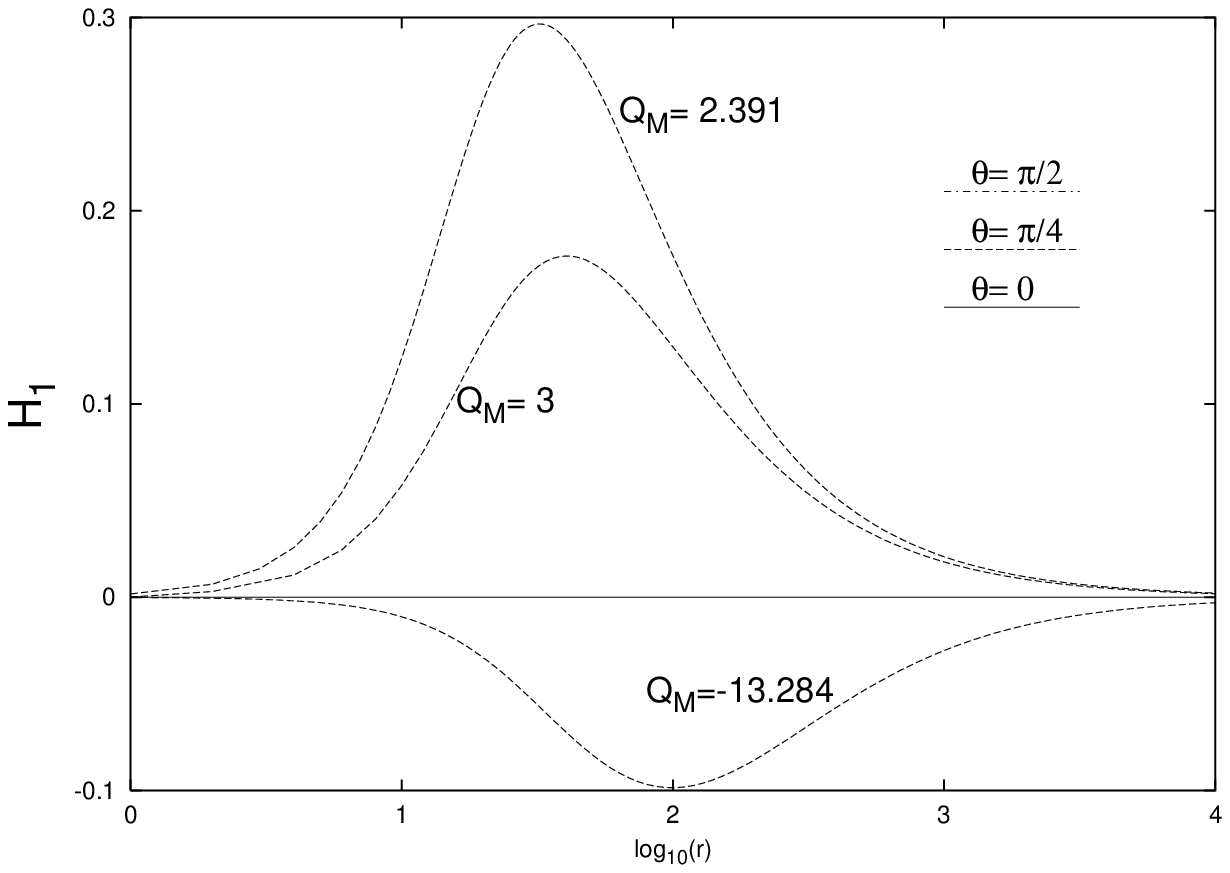,width=16cm}}
\end{picture}
\begin{center}
Figure 1a.
\end{center}

\newpage

\begin{picture}(16,16)
\centering
\put(-2,0){\epsfig{file=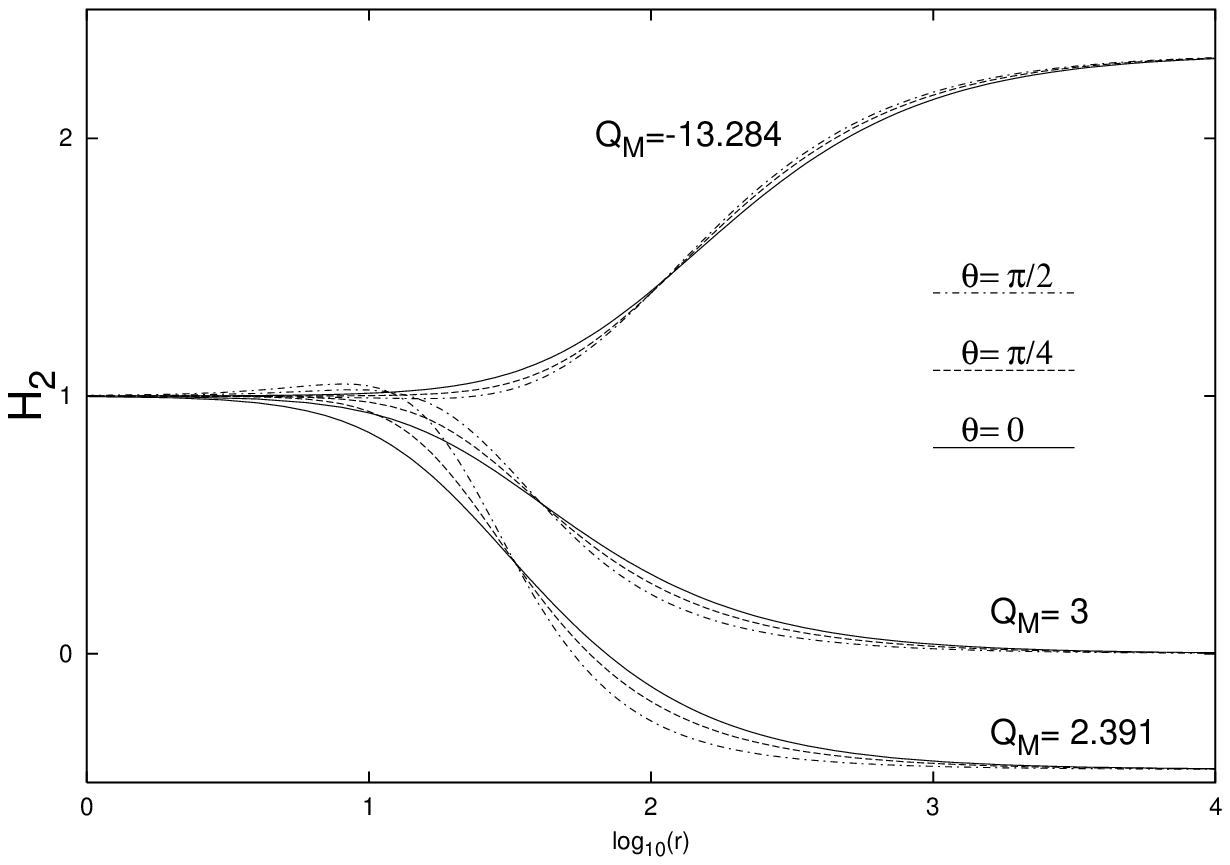,width=16cm}}
\end{picture}
\begin{center}
Figure 1b.
\end{center}

\newpage

\begin{picture}(16,16)
\centering
\put(-2,0){\epsfig{file=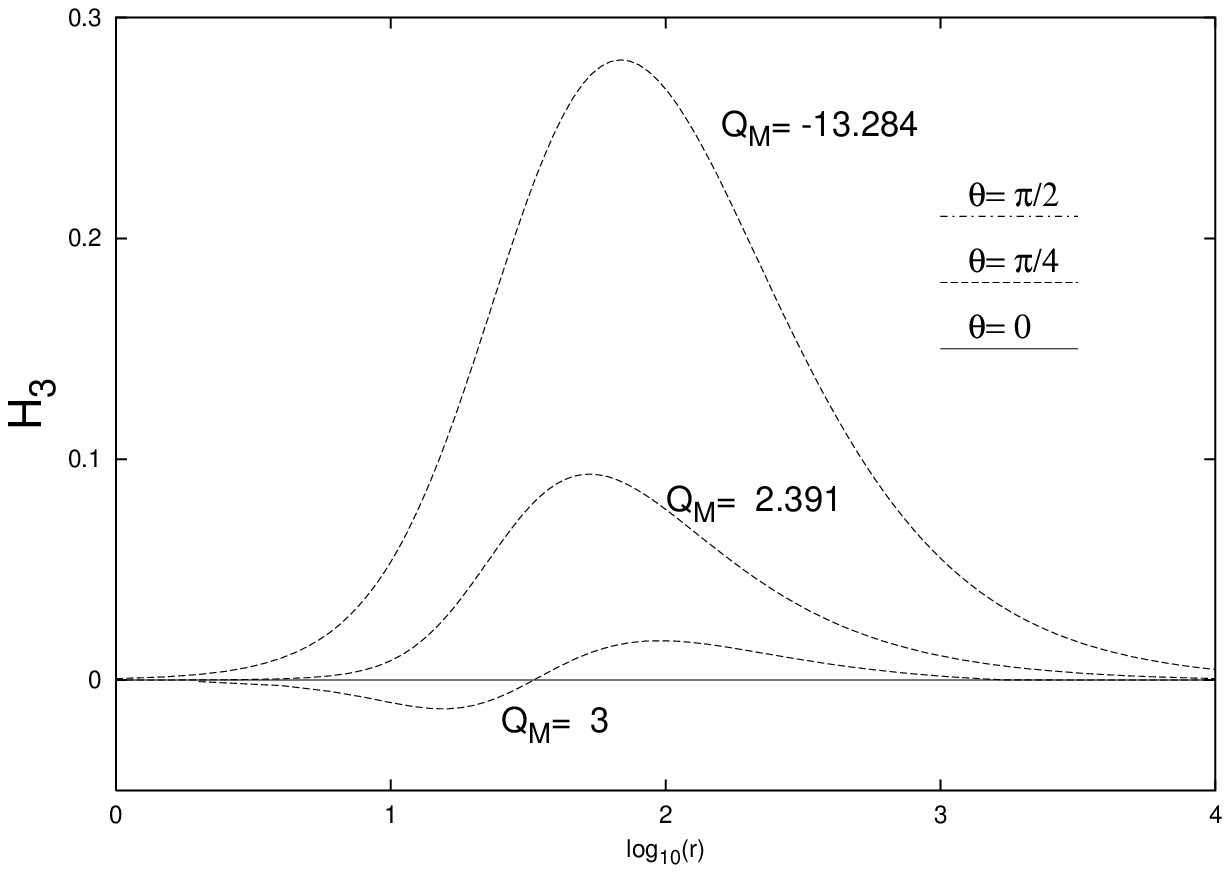,width=16cm}}
\end{picture}
\begin{center}
Figure 1c
\end{center}

\newpage

\begin{picture}(16,16)
\centering
\put(-2,0){\epsfig{file=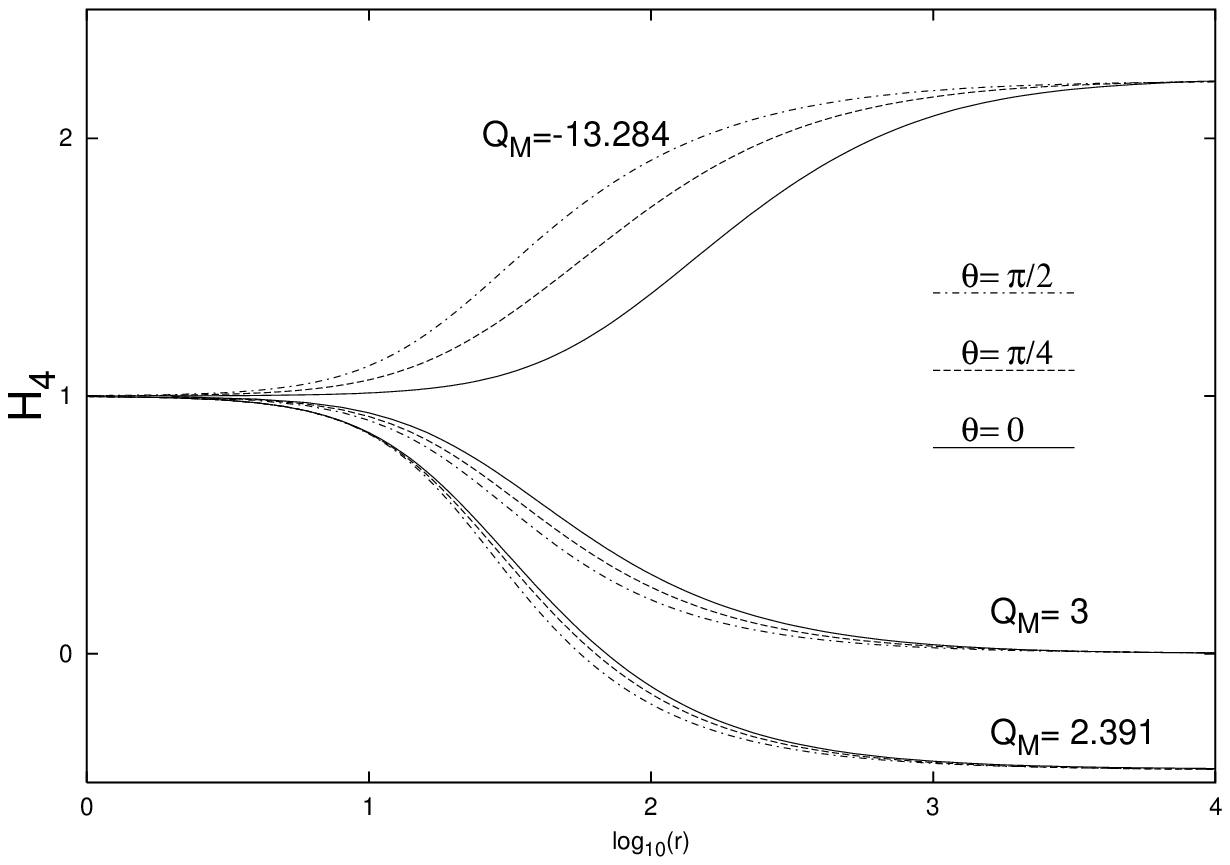,width=16cm}}
\end{picture}
\begin{center}
Figure 1d
\end{center}

\newpage

\begin{picture}(16,16)
\centering
\put(-2,0){\epsfig{file=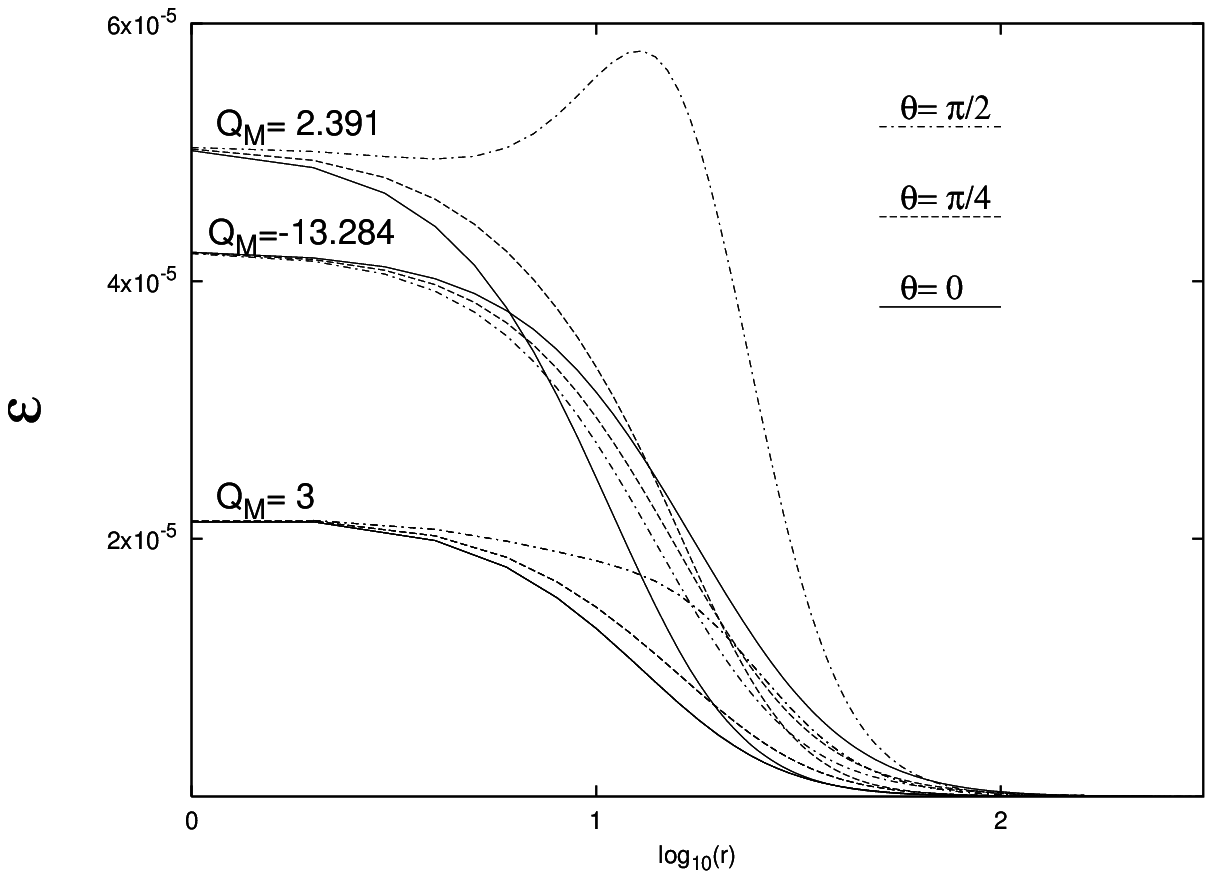,width=16cm}}
\end{picture}
\begin{center}
Figure 1e
\end{center}

\setlength{\unitlength}{1cm}

\begin{picture}(16,16)
\centering
\put(-2,0){\epsfig{file=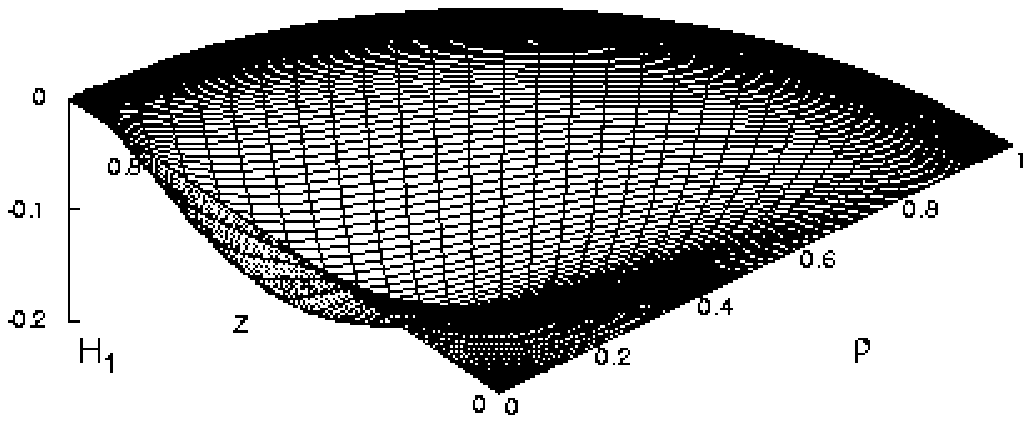,width=16cm}}
\end{picture}
\begin{center}
Figure 2a.
\end{center}

\newpage

\begin{picture}(16,16)
\centering
\put(-2,0){\epsfig{file=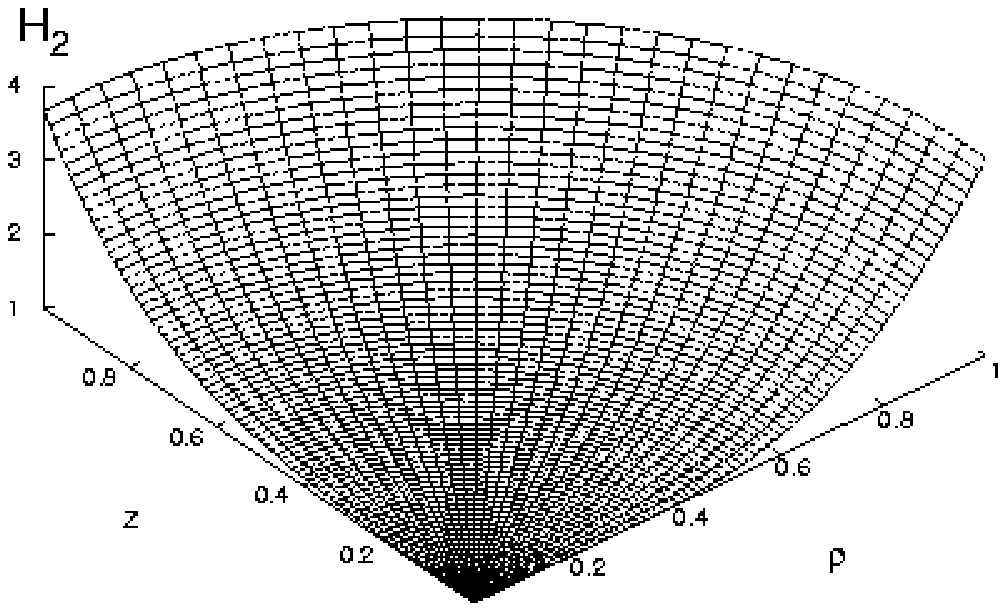,width=16cm}}
\end{picture}
\begin{center}
Figure 2b.
\end{center}

\newpage

\begin{picture}(16,16)
\centering
\put(-2,0){\epsfig{file=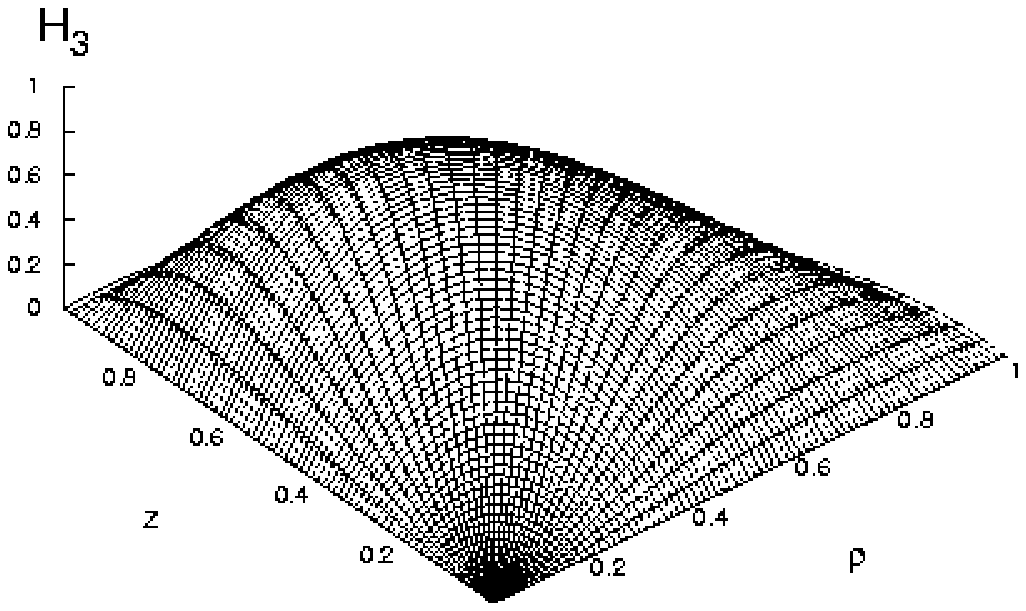,width=16cm}}
\end{picture}
\begin{center}
Figure 2c
\end{center}

\newpage

\begin{picture}(16,16)
\centering
\put(-2,0){\epsfig{file=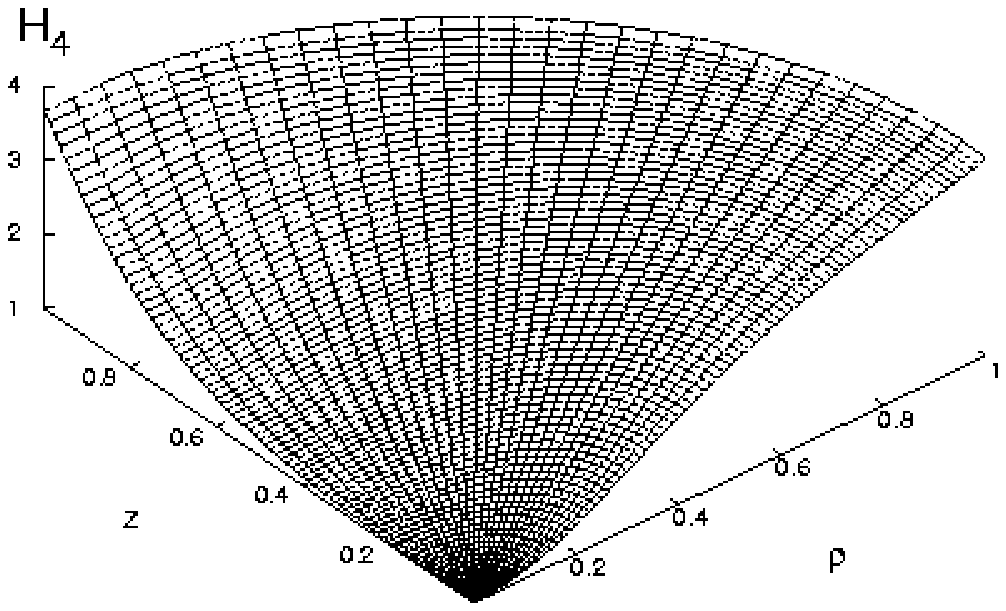,width=16cm}}
\end{picture}
\begin{center}
Figure 2d
\end{center}

\newpage

\begin{picture}(16,16)
\centering
\put(-2,0){\epsfig{file=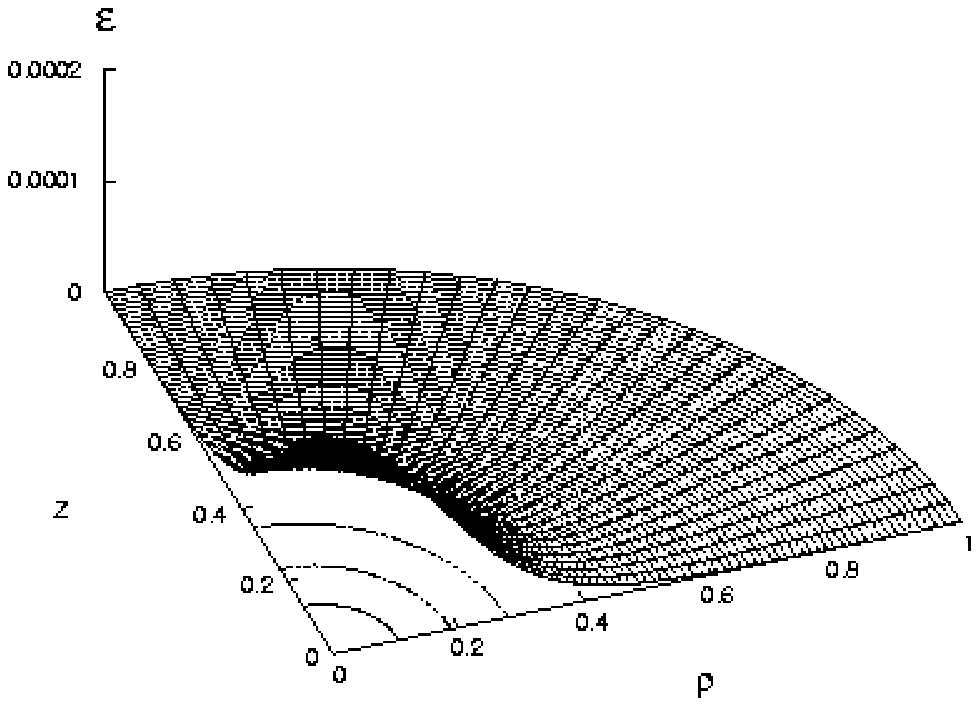,width=16cm}}
\end{picture}
\begin{center}
Figure 2e
\end{center}

\begin{picture}(16,16)
\centering
\put(-2,0){\epsfig{file=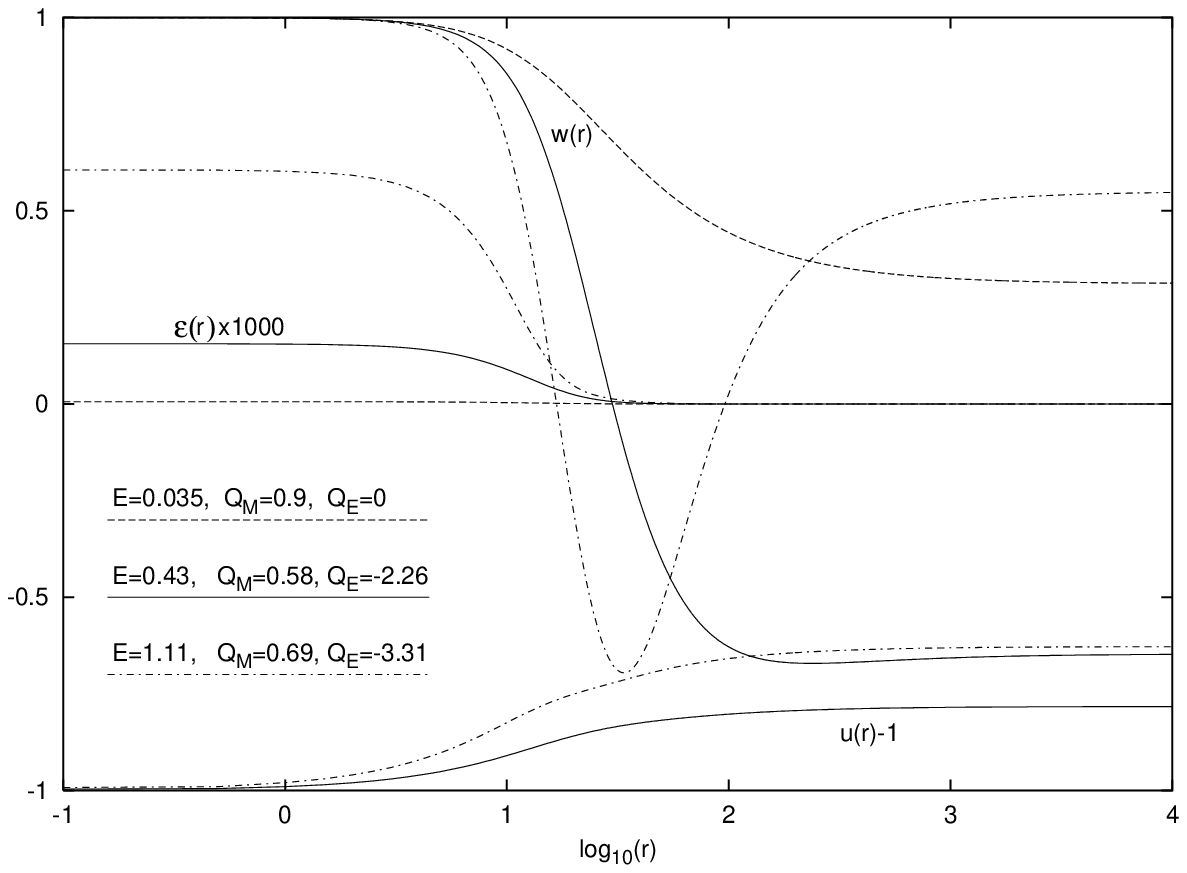,width=16cm}}
\end{picture}
\begin{center}
Figure 3
\end{center}
\newpage

\newpage
\begin{picture}(16,16)
\centering
\put(-2,0){\epsfig{file=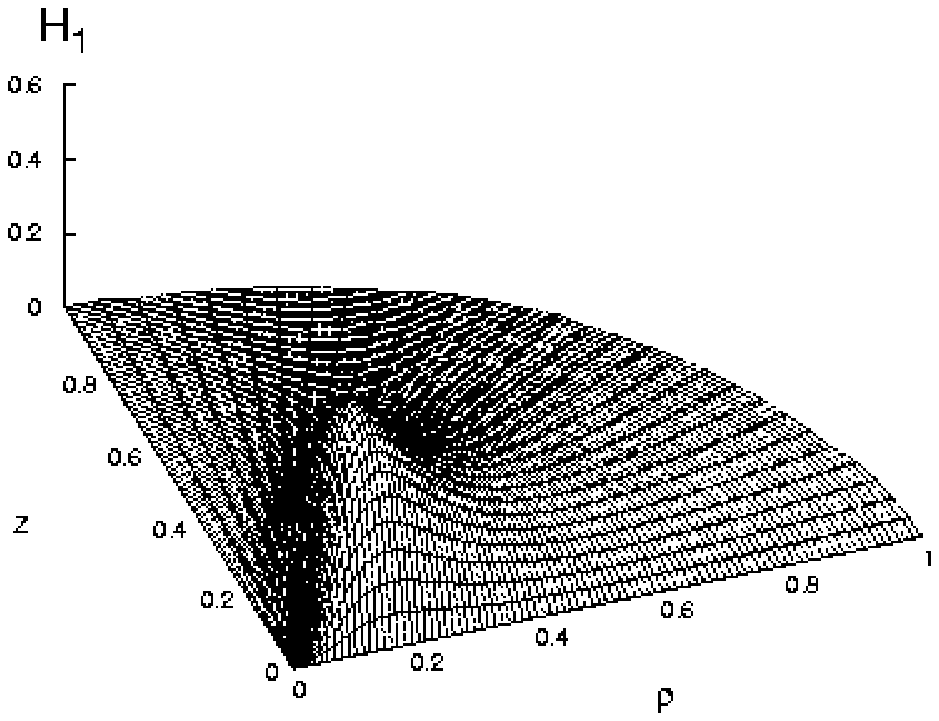,width=16cm}}
\end{picture}
\begin{center}
Figure 4a.
\end{center}

\newpage

\begin{picture}(16,16)
\centering
\put(-2,0){\epsfig{file=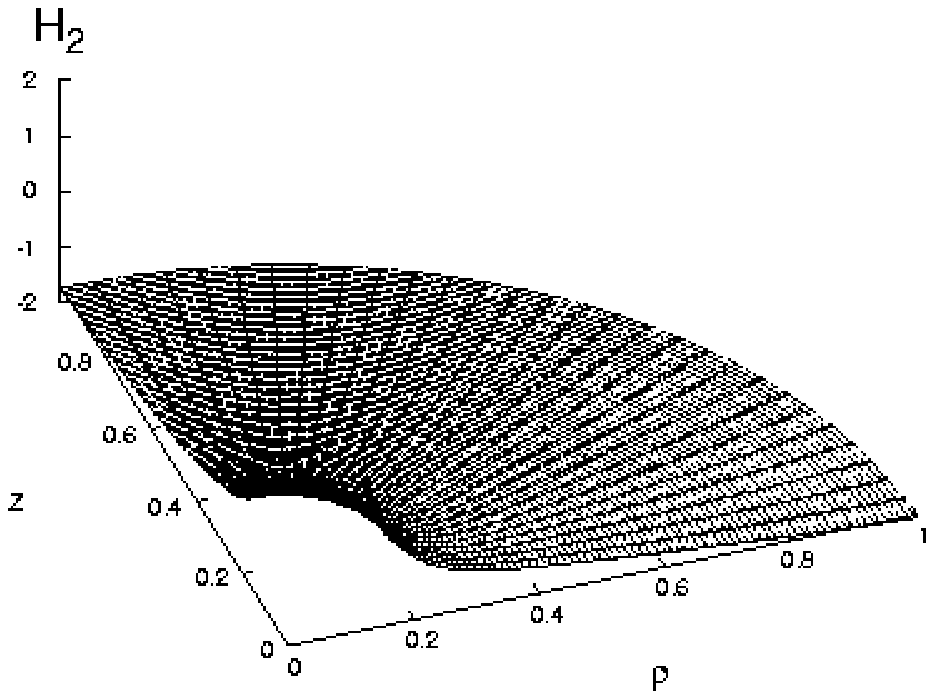,width=16cm}}
\end{picture}
\begin{center}
Figure 4b
\end{center}

\newpage

\begin{picture}(16,16)
\centering
\put(-2,0){\epsfig{file=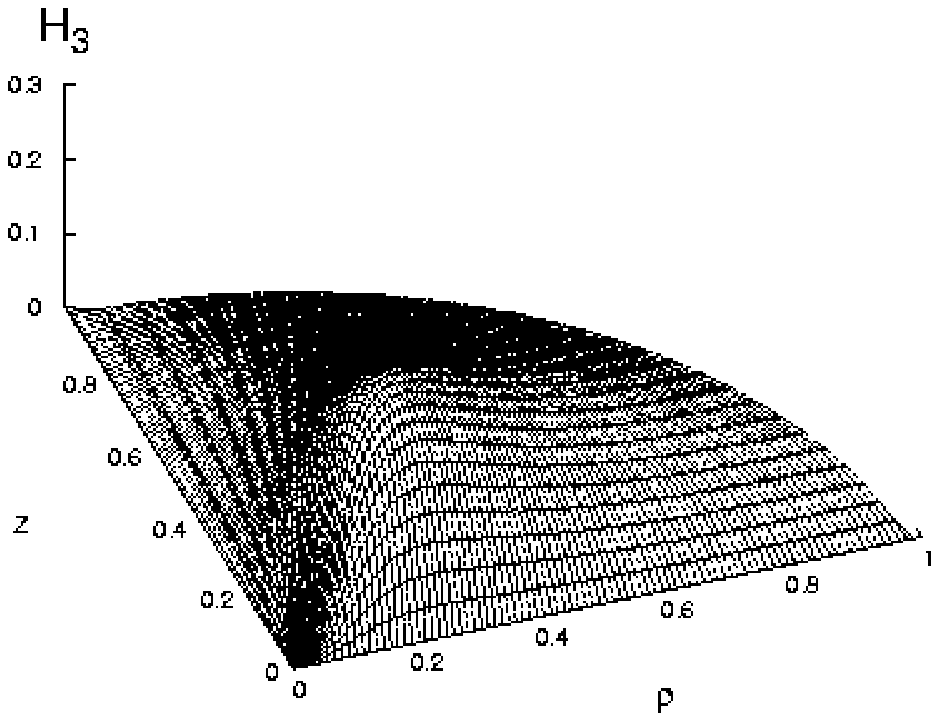,width=16cm}}
\end{picture}
\begin{center}
Figure 4c
\end{center}

\newpage

\begin{picture}(16,16)
\centering
\put(-2,0){\epsfig{file=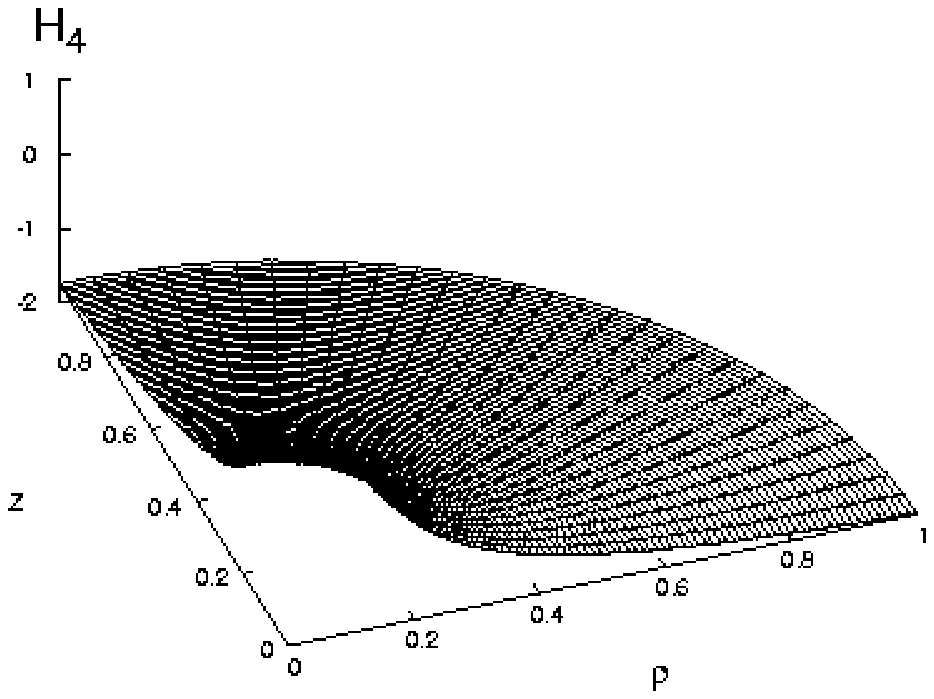,width=16cm}}
\end{picture}
\begin{center}
Figure 4d
\end{center}

\newpage

\begin{picture}(16,16)
\centering
\put(-2,0){\epsfig{file=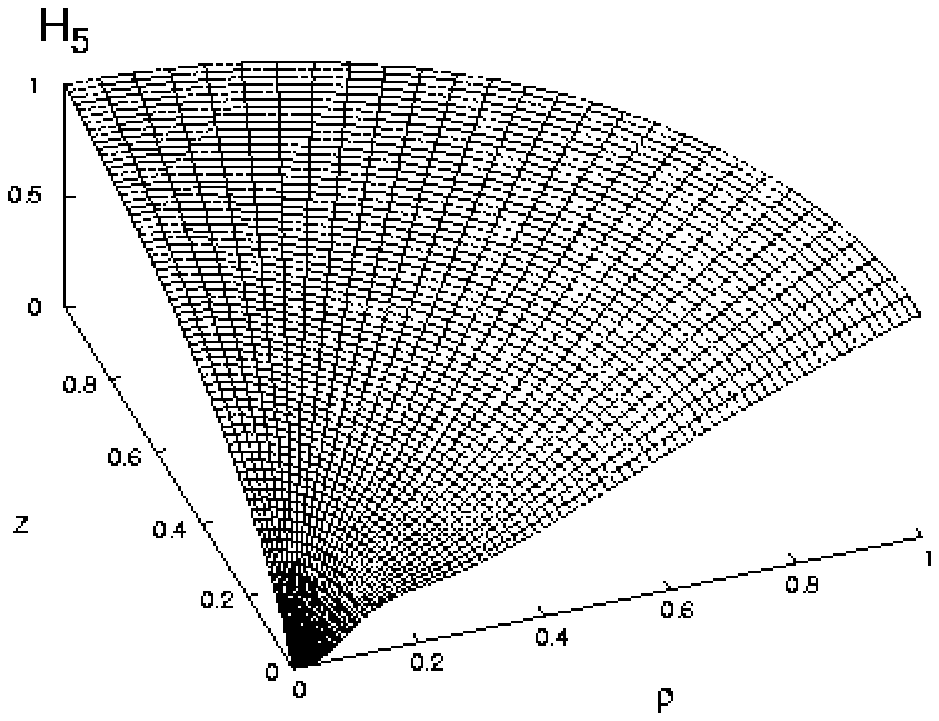,width=16cm}}
\end{picture}
\begin{center}
Figure 4e
\end{center}

\newpage

\begin{picture}(16,16)
\centering
\put(-2,0){\epsfig{file=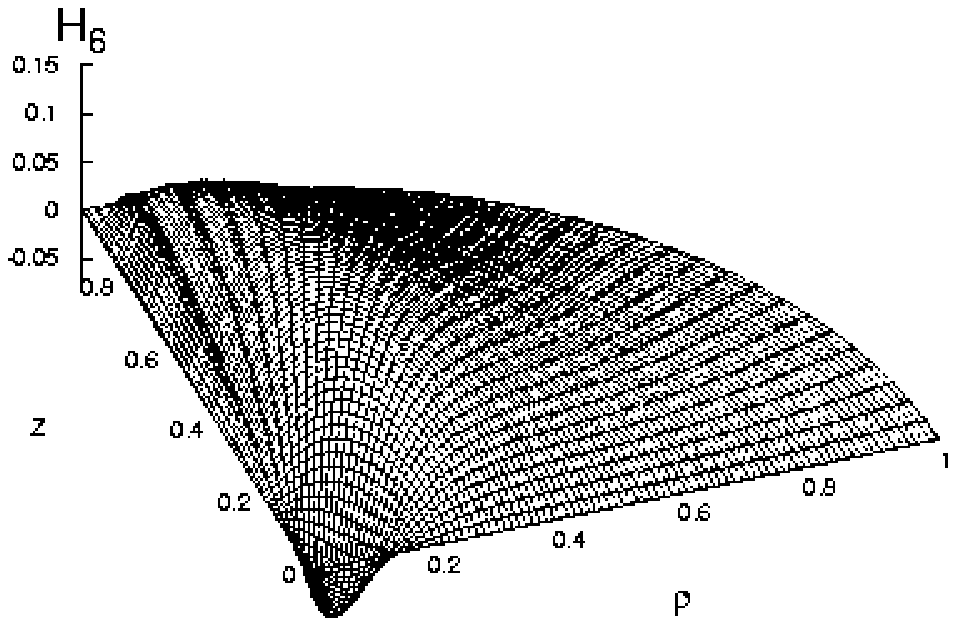,width=16cm}}
\end{picture}
\begin{center}
Figure 4f
\end{center}

\newpage

\begin{picture}(16,16)
\centering
\put(-2,0){\epsfig{file=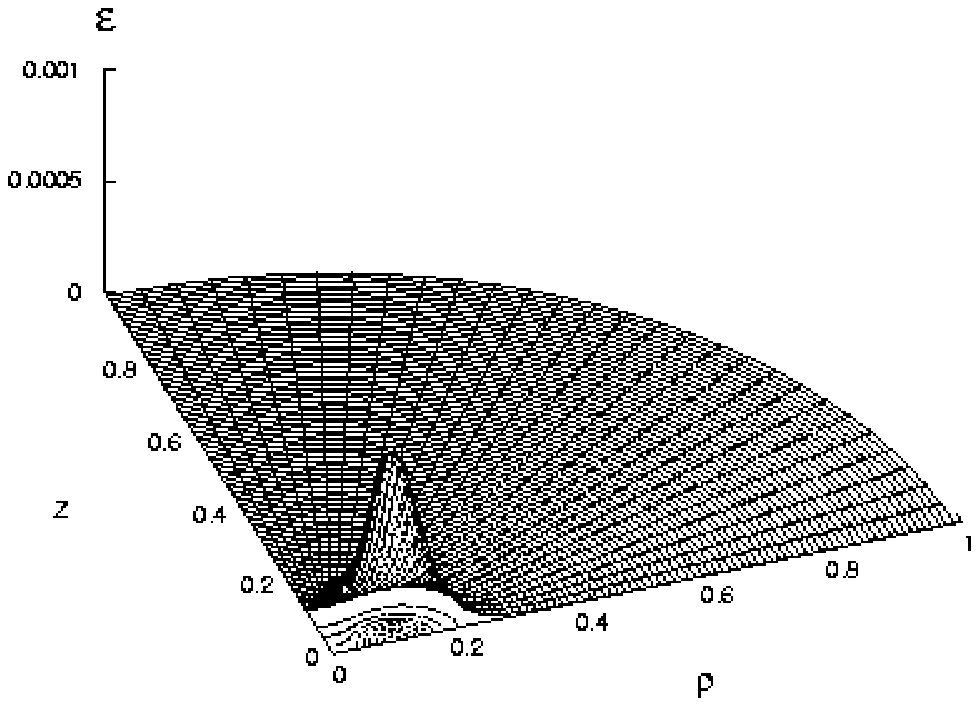,width=16cm}}
\end{picture}
\begin{center}
Figure 4g.
\end{center}

\newpage
\begin{picture}(16,16)
\centering
\put(-2,0){\epsfig{file=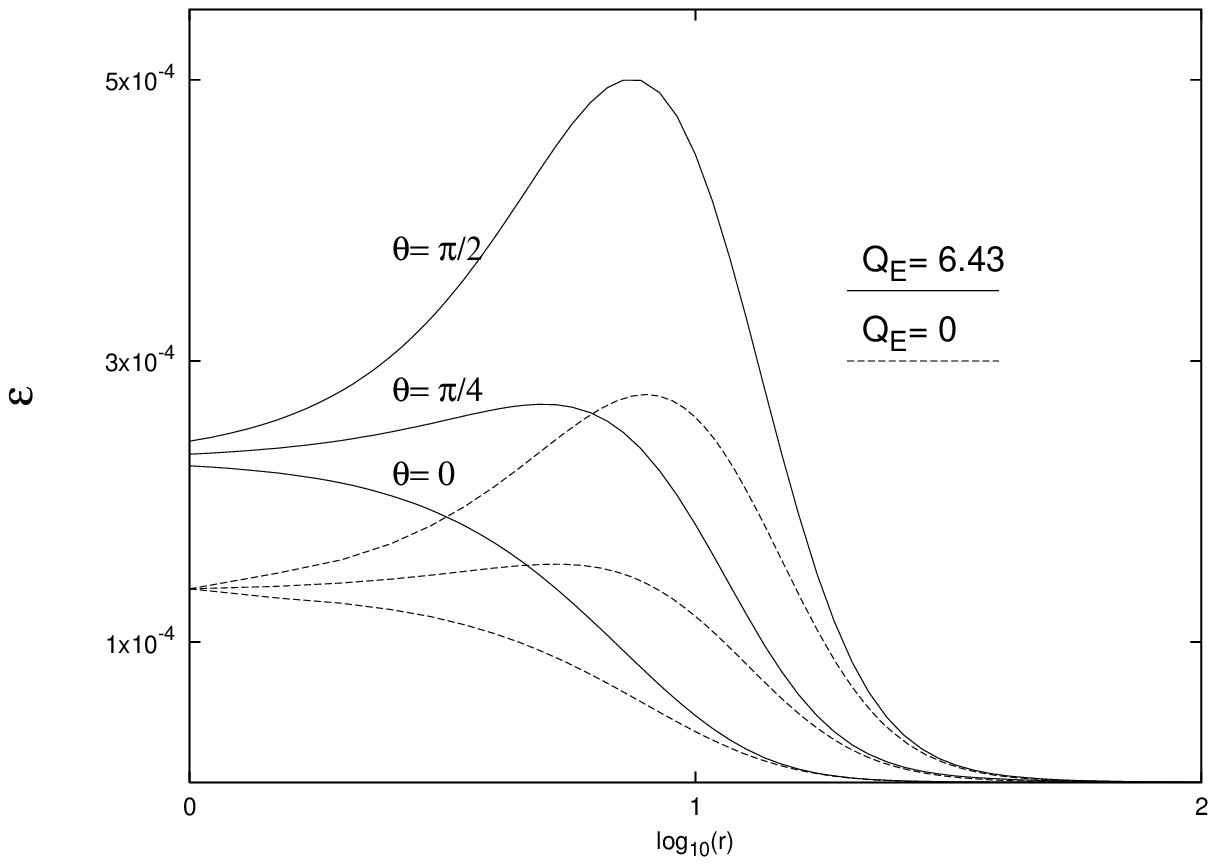,width=16cm}}
\end{picture}
\begin{center}
Figure 5
\end{center}

\newpage
\begin{picture}(16,16)
\centering
\put(-2,0){\epsfig{file=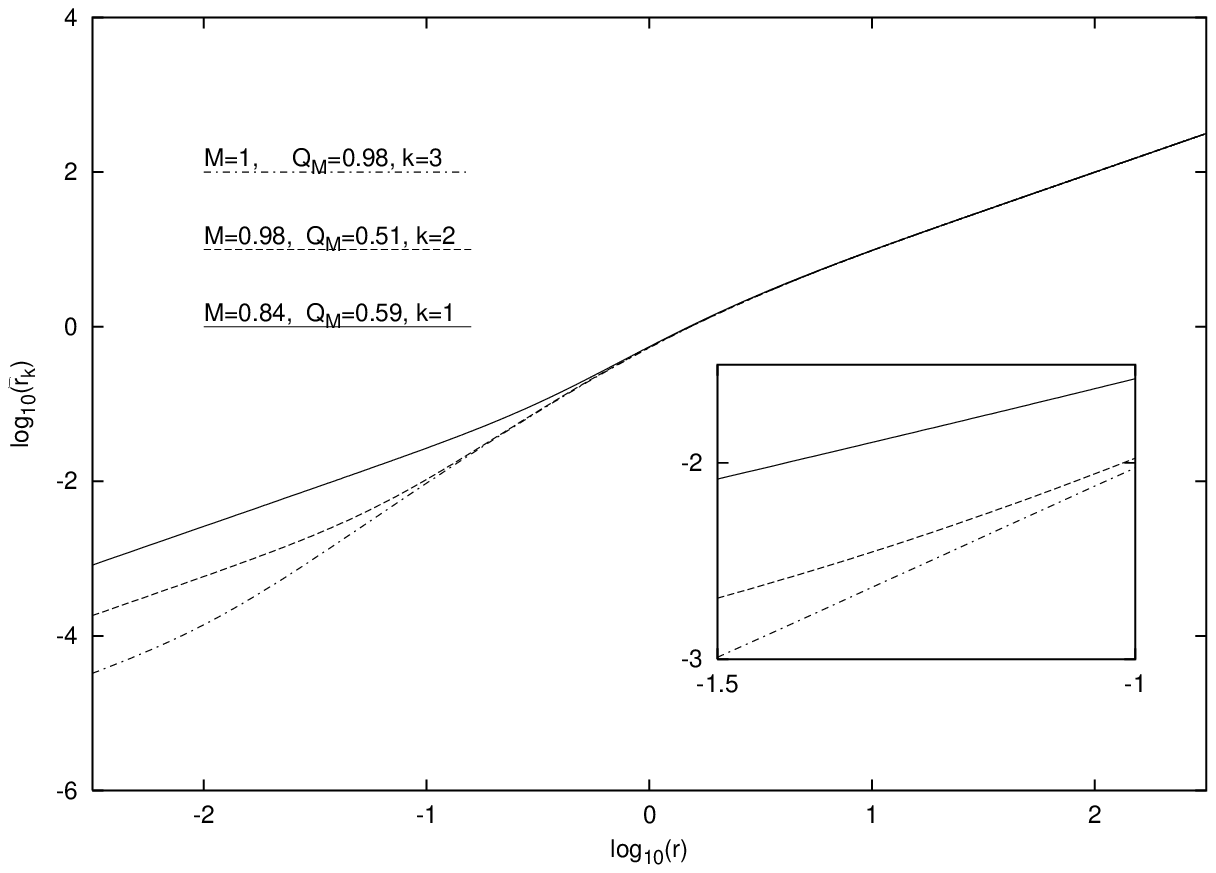,width=16cm}}
\end{picture}
\begin{center}
Figure 6a
\end{center}

\newpage
\begin{picture}(16,16)
\centering
\put(-2,0){\epsfig{file=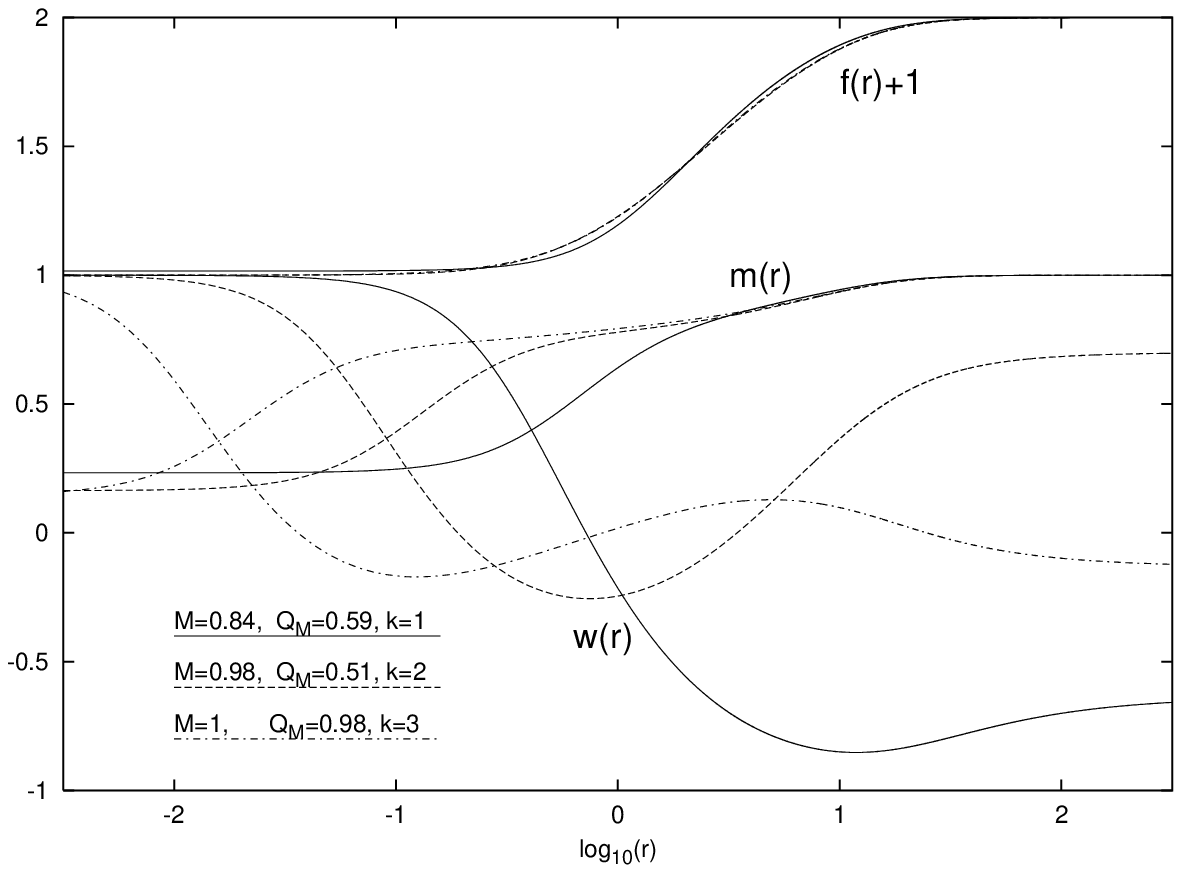,width=16cm}}
\end{picture}
\begin{center}
Figure 6b
\end{center}

\newpage
\begin{picture}(16,16)
\centering
\put(-2,0){\epsfig{file=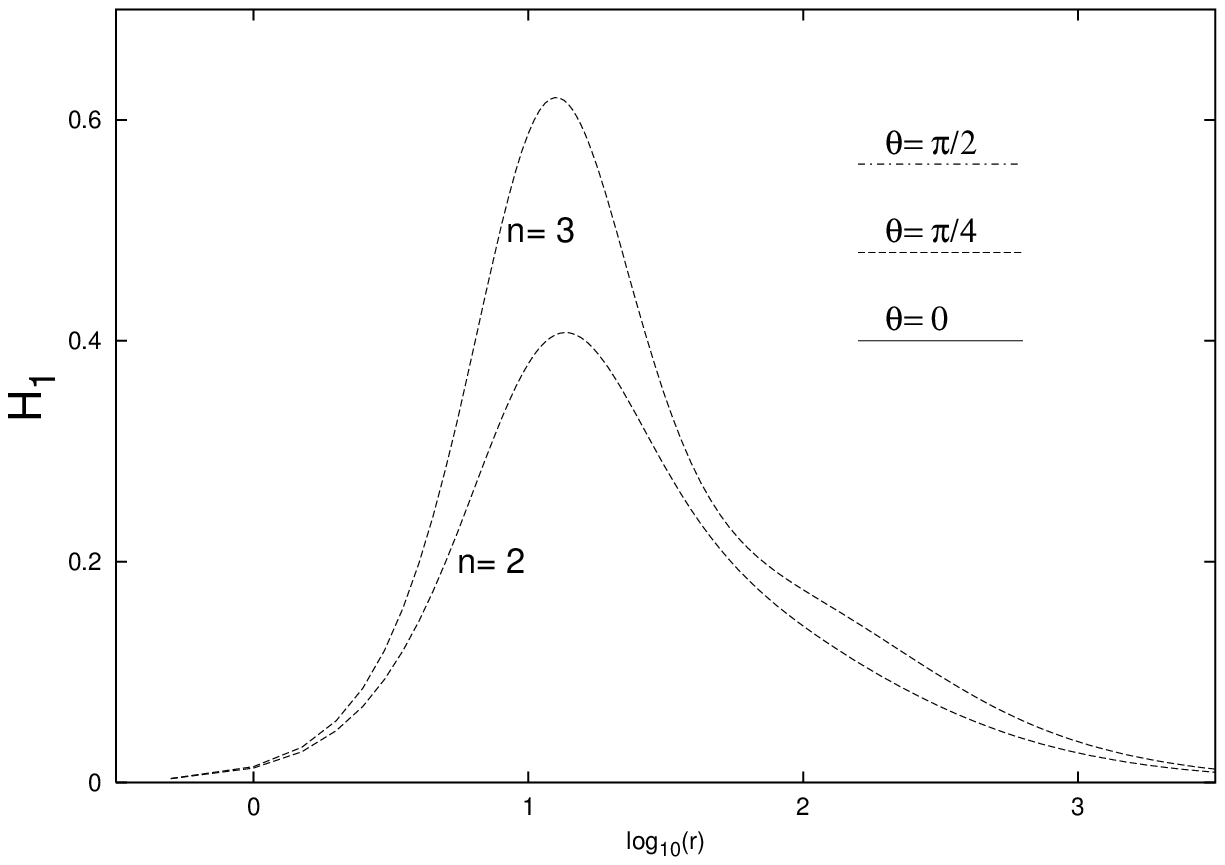,width=16cm}}
\end{picture}
\begin{center}
Figure 7a
\end{center}

\newpage
\begin{picture}(16,16)
\centering
\put(-2,0){\epsfig{file=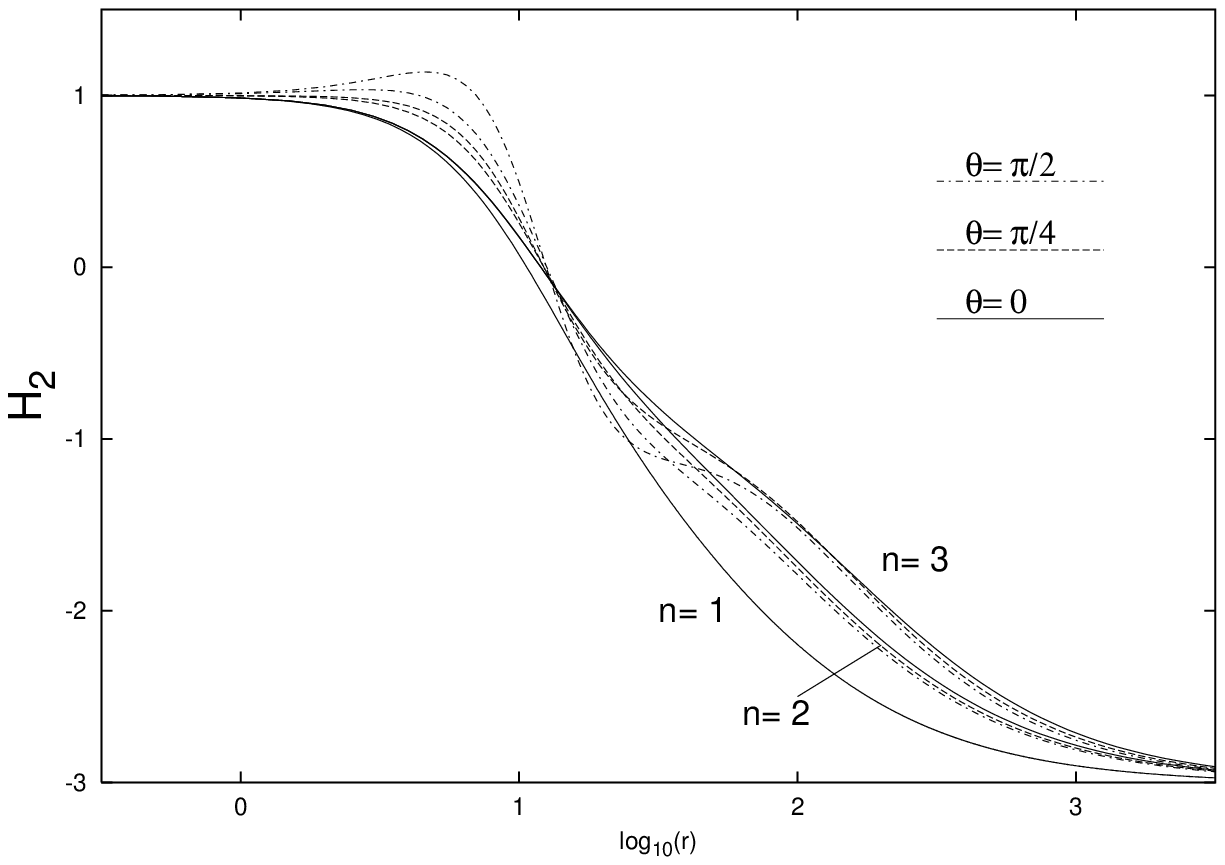,width=16cm}}
\end{picture}
\begin{center}
Figure 7b
\end{center}

\newpage
\begin{picture}(16,16)
\centering
\put(-2,0){\epsfig{file=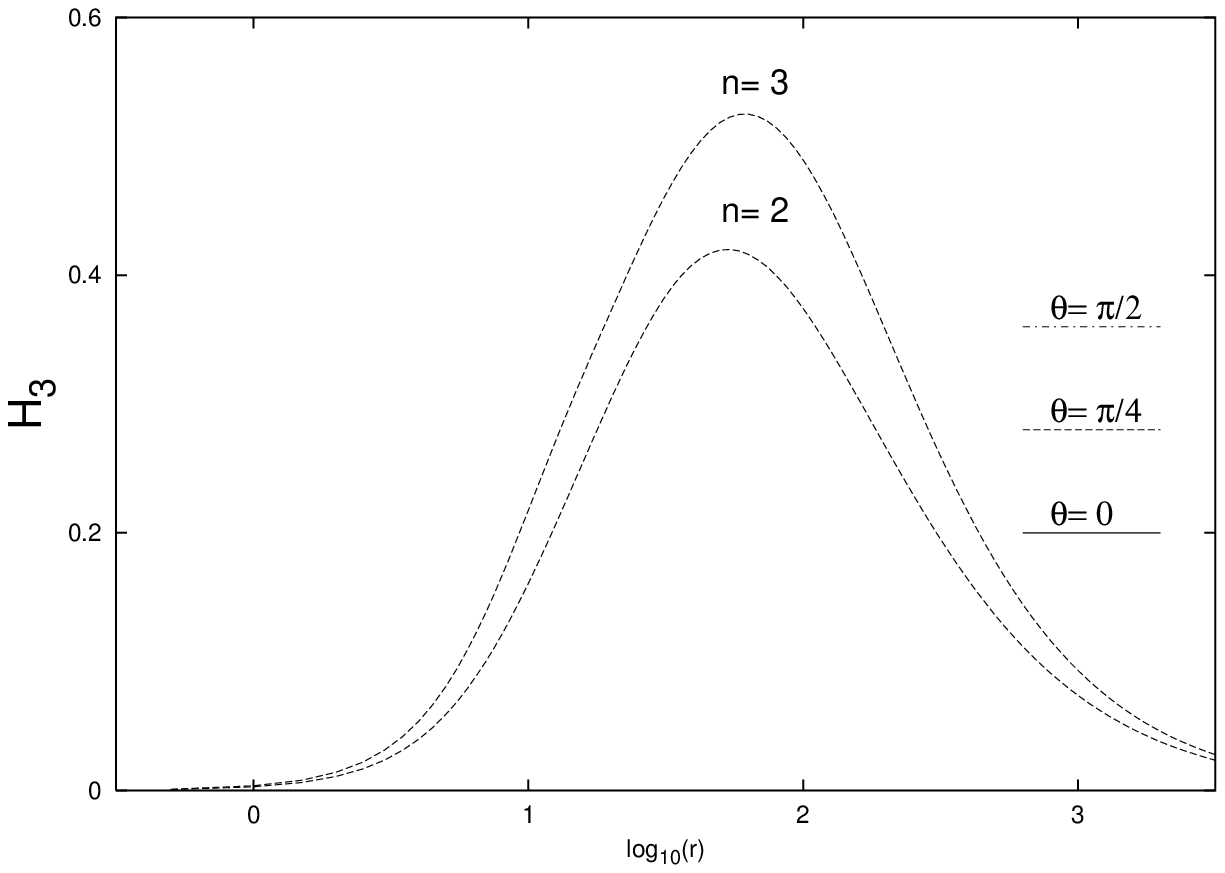,width=16cm}}
\end{picture}
\begin{center}
Figure 7c
\end{center}

\newpage
\begin{picture}(16,16)
\centering
\put(-2,0){\epsfig{file=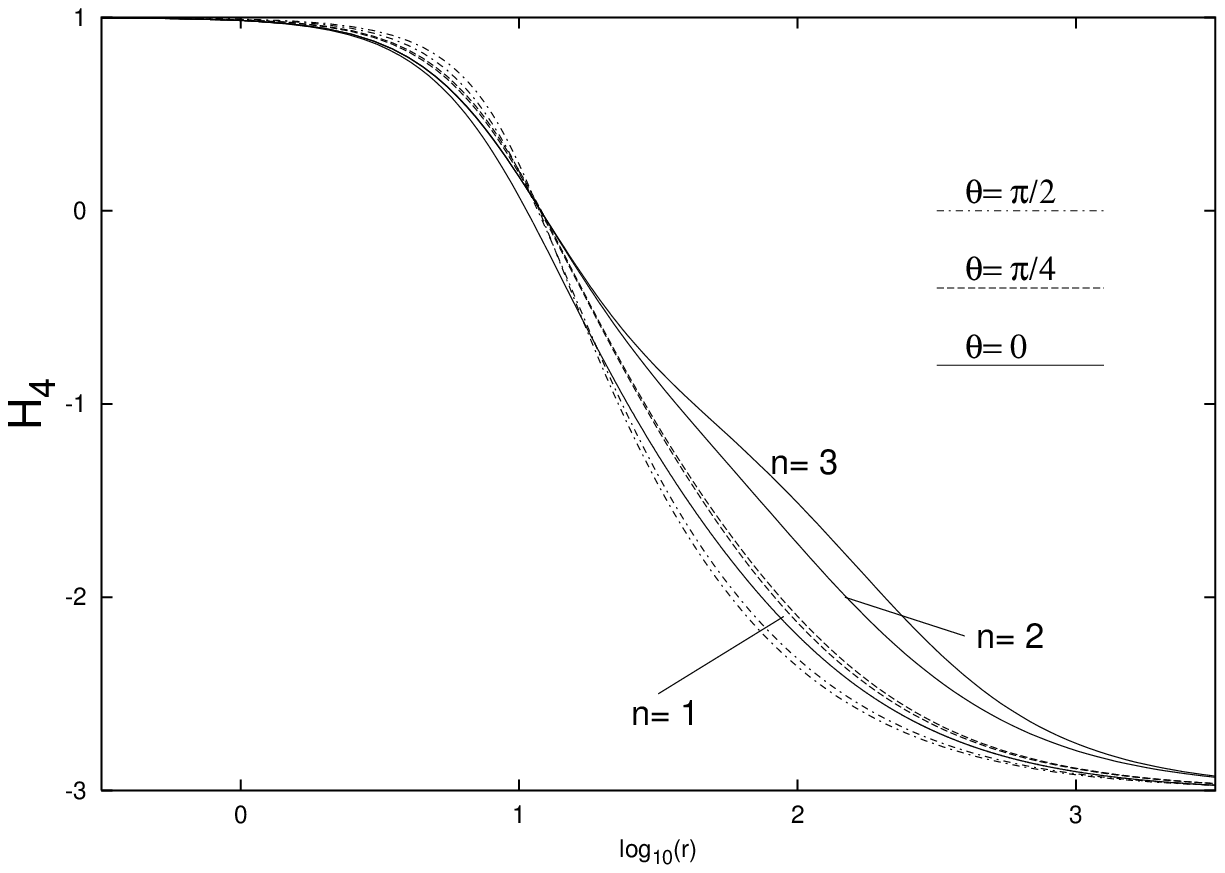,width=16cm}}
\end{picture}
\begin{center}
Figure 7d
\end{center}

\newpage
\begin{picture}(16,16)
\centering
\put(-2,0){\epsfig{file=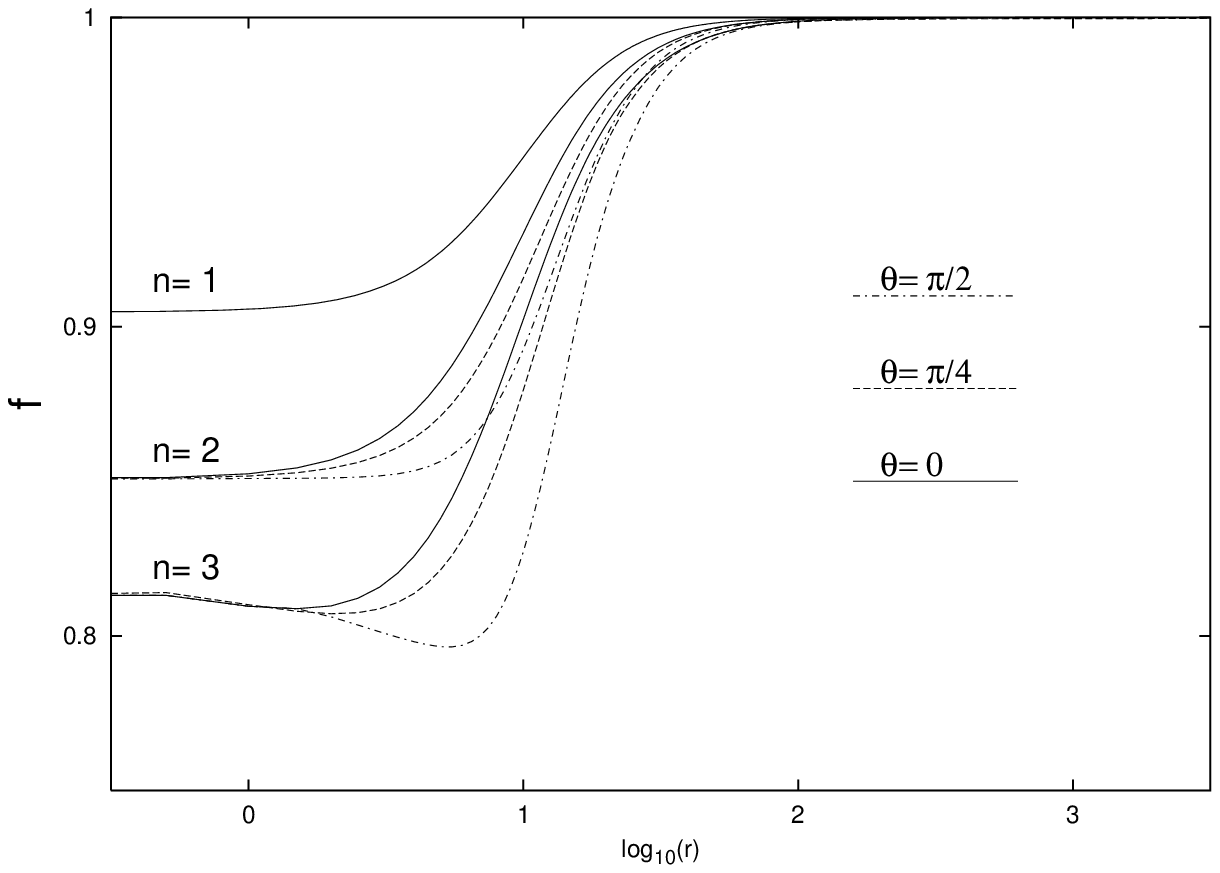,width=16cm}}
\end{picture}
\begin{center}
Figure 7e
\end{center}

\newpage
\begin{picture}(16,16)
\centering
\put(-2,0){\epsfig{file=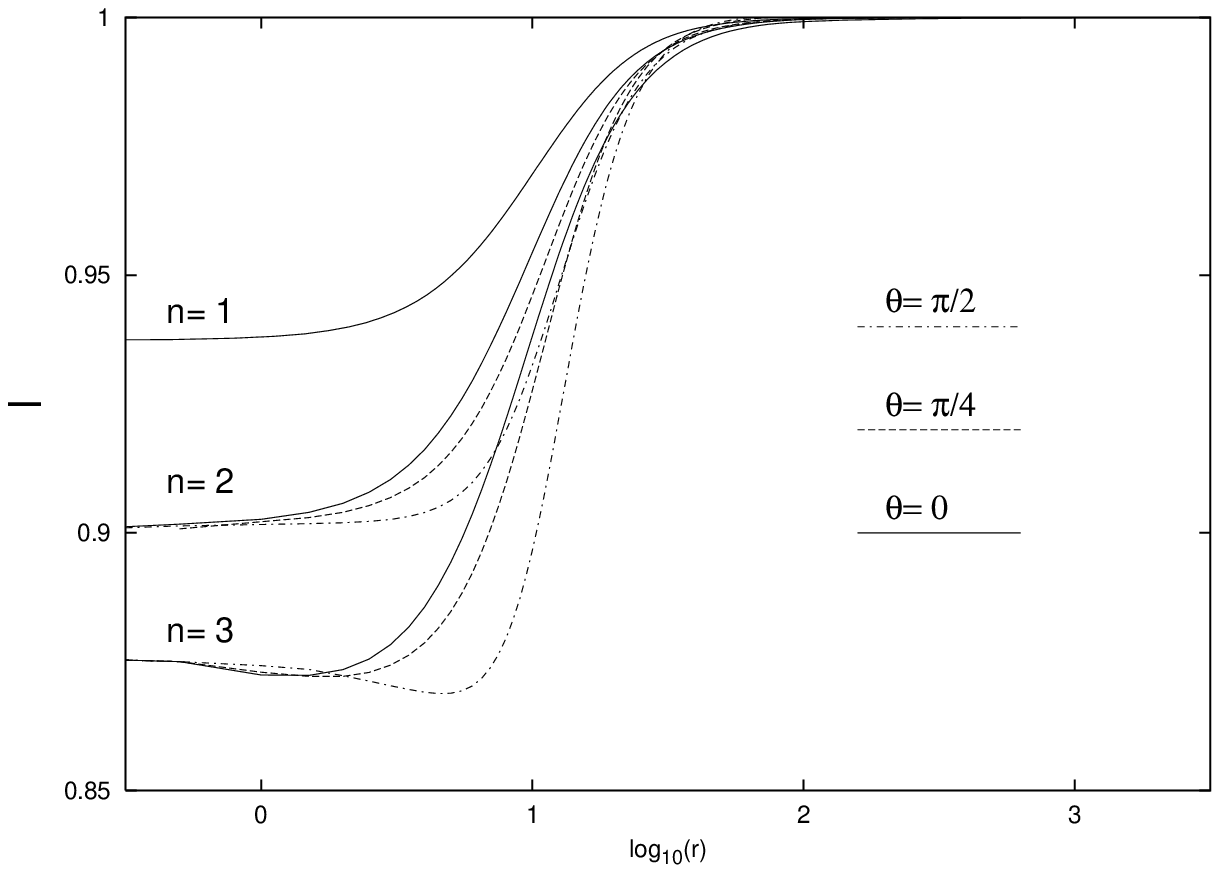,width=16cm}}
\end{picture}
\begin{center}
Figure 7f
\end{center}

\newpage
\begin{picture}(16,16)
\centering
\put(-2,0){\epsfig{file=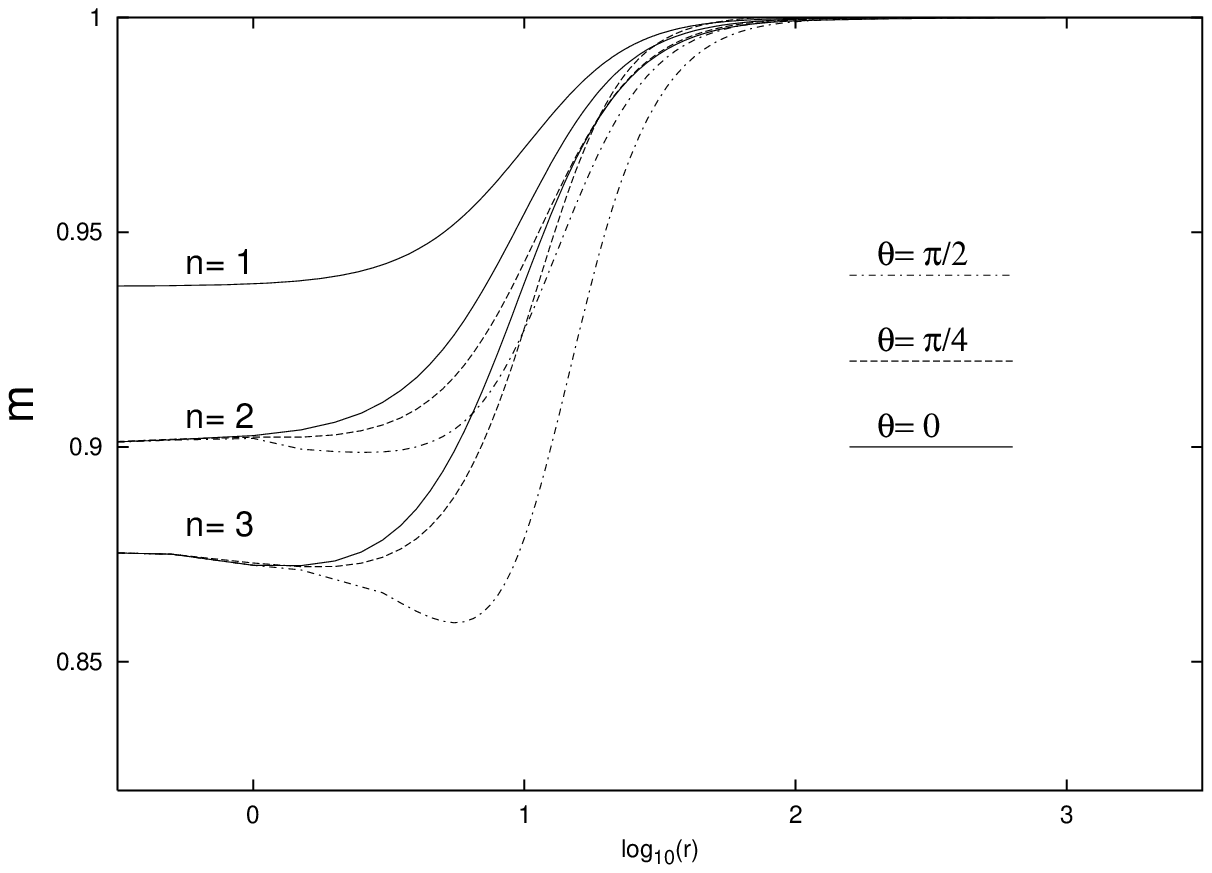,width=16cm}}
\end{picture}
\begin{center}
Figure 7g
\end{center}

\newpage
\begin{picture}(16,16)
\centering
\put(-2,0){\epsfig{file=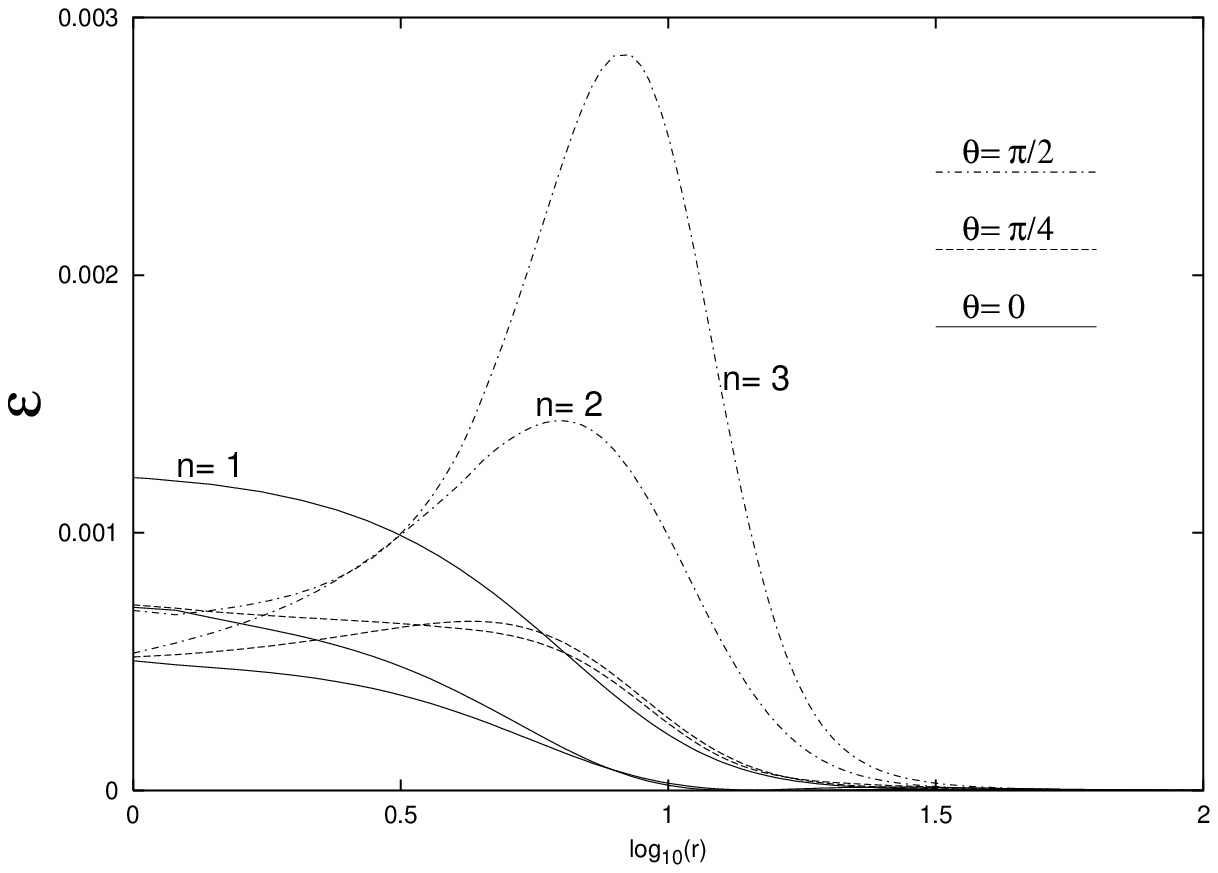,width=16cm}}
\end{picture}
\begin{center}
Figure 7h
\end{center}

\newpage
\begin{picture}(16,16)
\centering
\put(-2,0){\epsfig{file=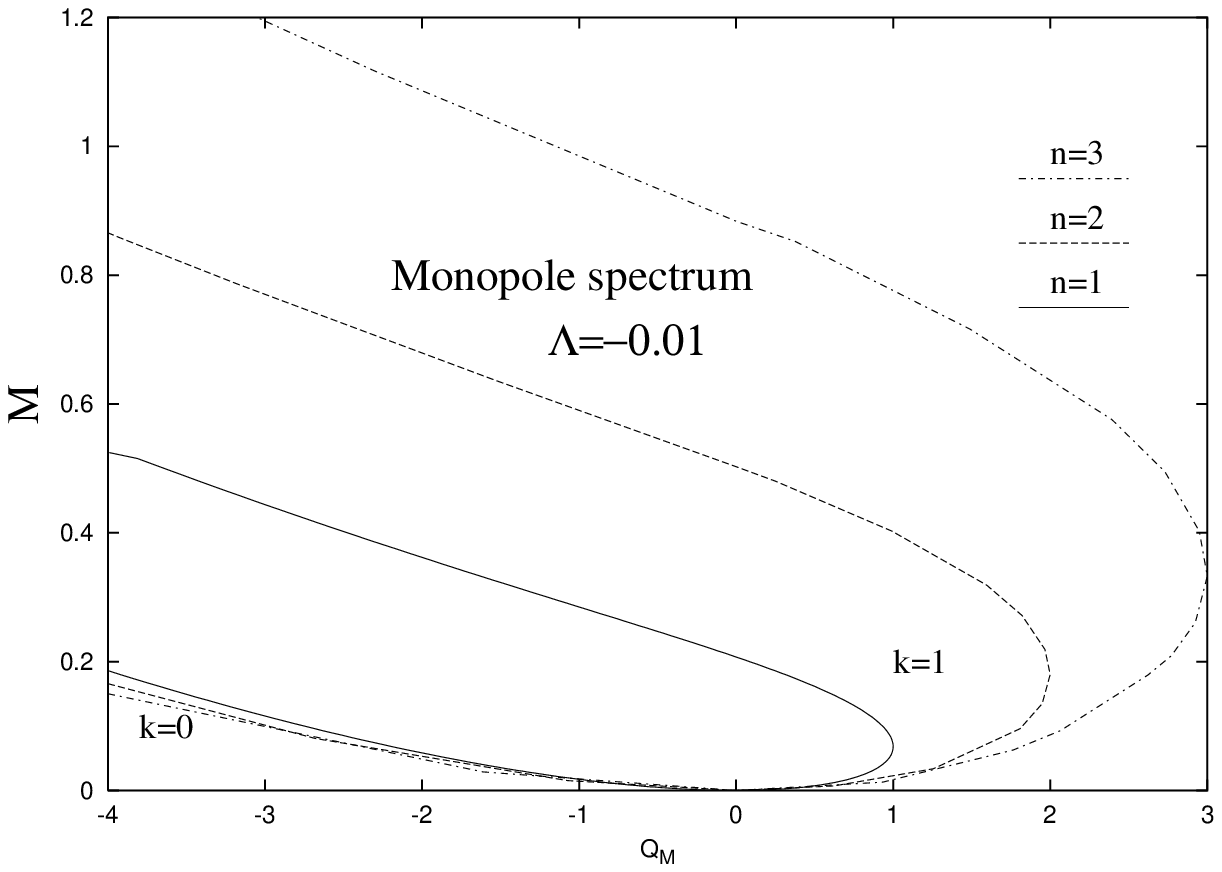,width=16cm}}
\end{picture}
\begin{center}
Figure 8
\end{center}

\newpage
\begin{picture}(16,16)
\centering
\put(-2,0){\epsfig{file=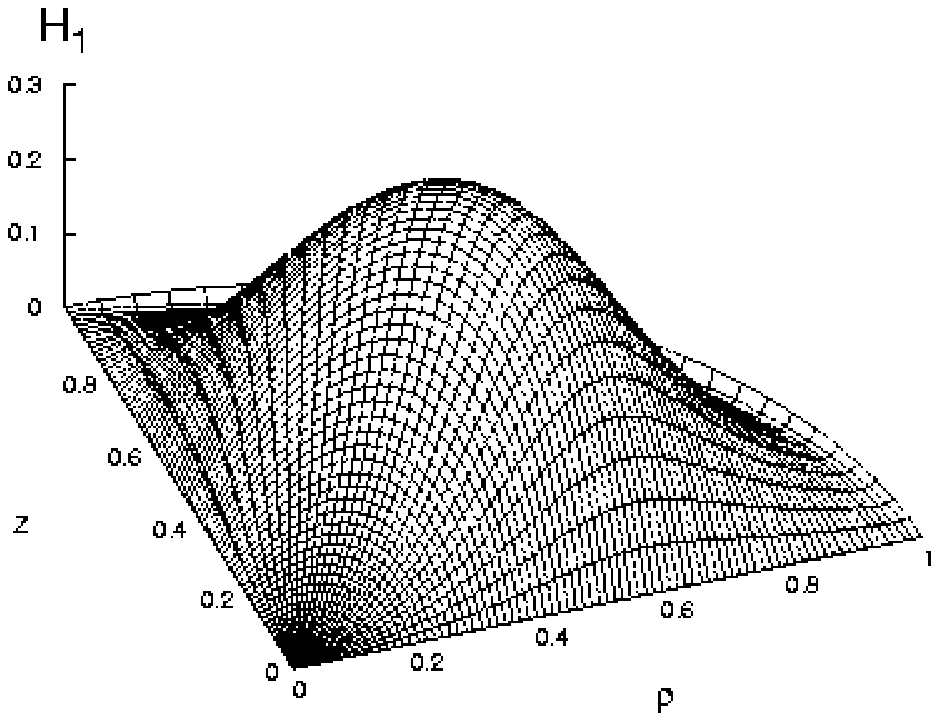,width=16cm}}
\end{picture}
\begin{center}
Figure 9a
\end{center}

\newpage
\begin{picture}(16,16)
\centering
\put(-2,0){\epsfig{file=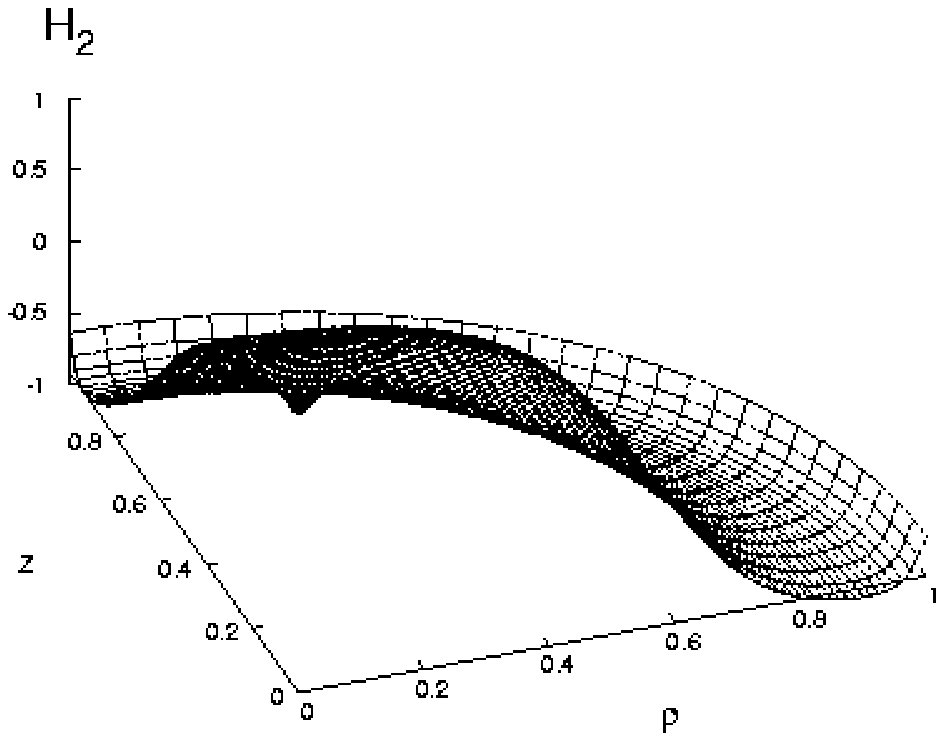,width=16cm}}
\end{picture}
\begin{center}
Figure 9b
\end{center}

\newpage
\begin{picture}(16,16)
\centering
\put(-2,0){\epsfig{file=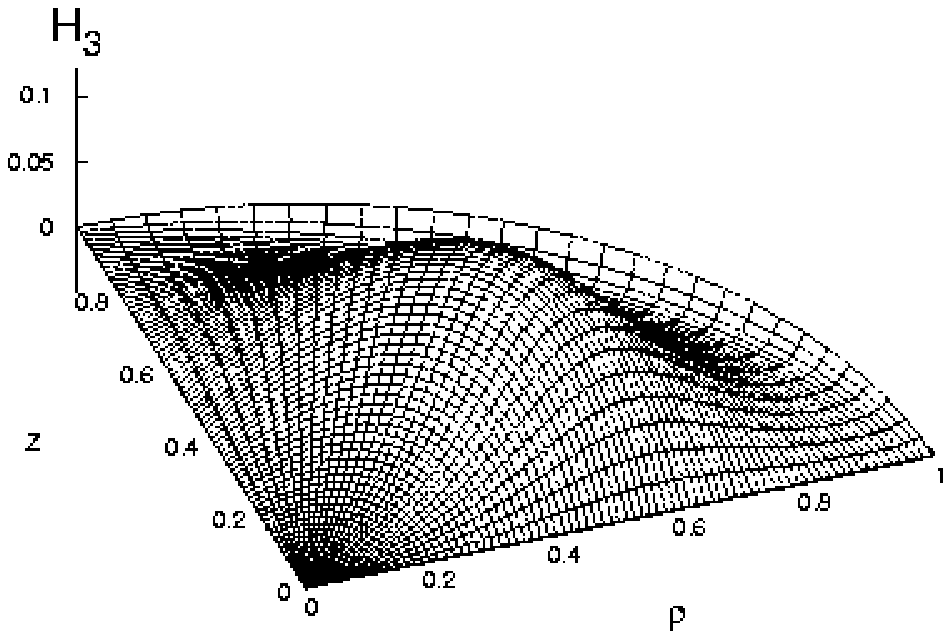,width=16cm}}
\end{picture}
\begin{center}
Figure 9c
\end{center}

\newpage
\begin{picture}(16,16)
\centering
\put(-2,0){\epsfig{file=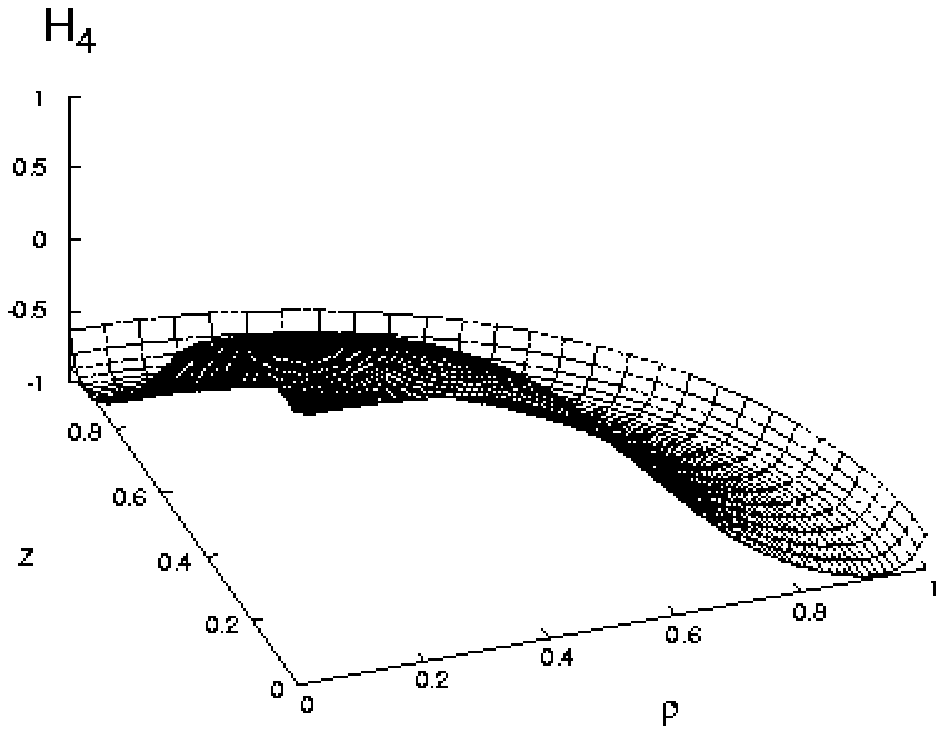,width=16cm}}
\end{picture}
\begin{center}
Figure 9d
\end{center}

\newpage
\begin{picture}(16,16)
\centering
\put(-2,0){\epsfig{file=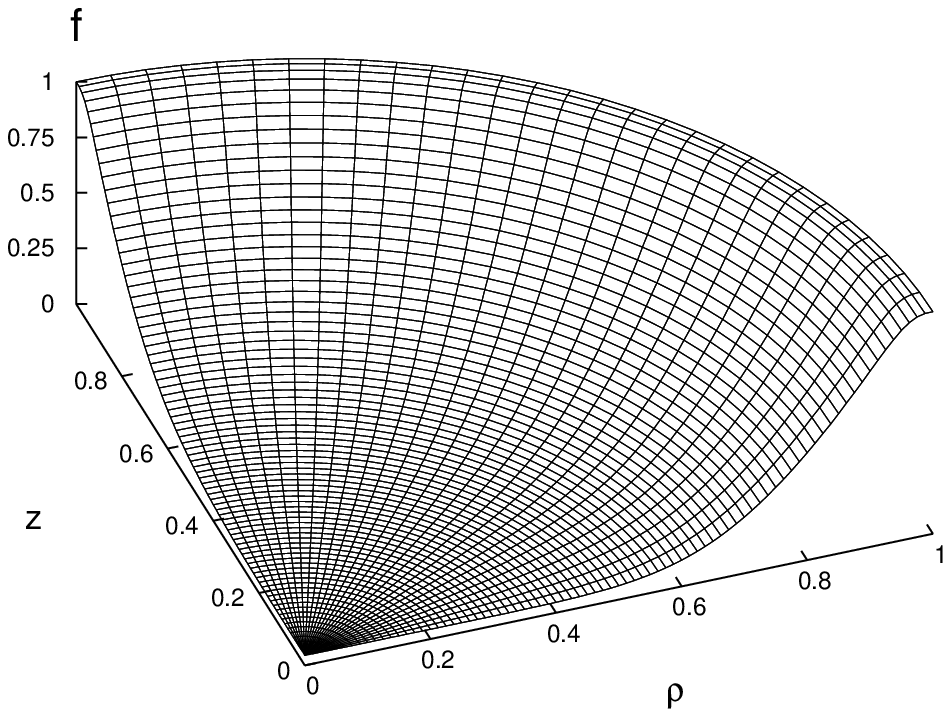,width=16cm}}
\end{picture}
\begin{center}
Figure 9e
\end{center}

\newpage
\begin{picture}(16,16)
\centering
\put(-2,0){\epsfig{file=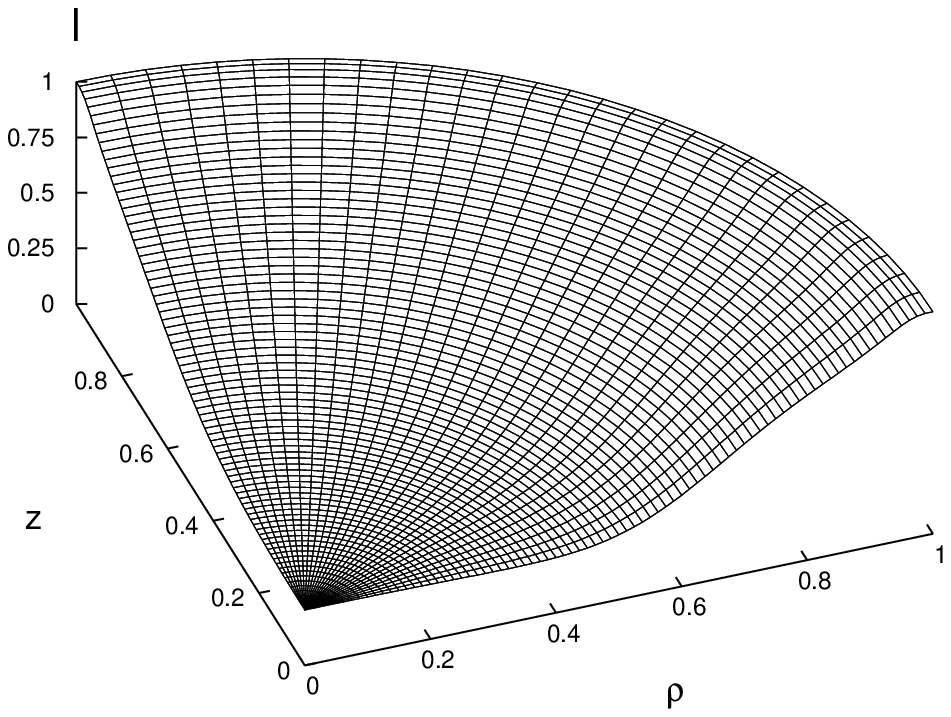,width=16cm}}
\end{picture}
\begin{center}
Figure 9f
\end{center}

\newpage
\begin{picture}(16,16)
\centering
\put(-2,0){\epsfig{file=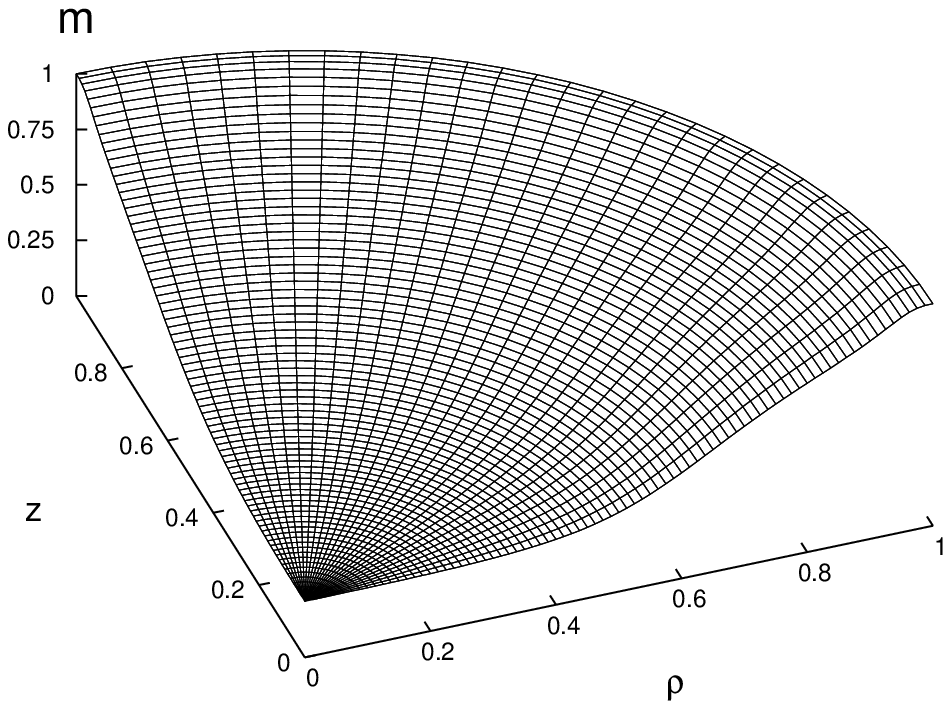,width=16cm}}
\end{picture}
\begin{center}
Figure 9g
\end{center}

\newpage
\begin{picture}(16,16)
\centering
\put(-2,0){\epsfig{file=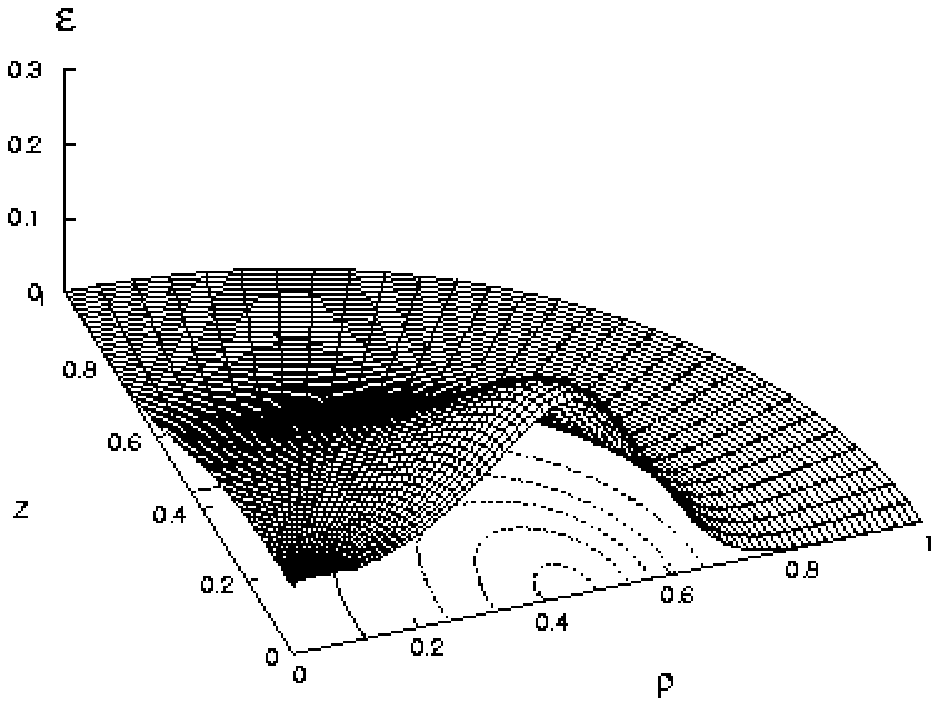,width=16cm}}
\end{picture}
\begin{center}
Figure 9h
\end{center}

\end{document}